\documentclass[fleqn,usenatbib,usedcolumn]{mnras}
\usepackage{graphicx,graphics,amsmath,amssymb,bm}
\usepackage{psfrag,epsfig,hyperref,tabularx,color,colortbl}
\usepackage[T1]{fontenc}
\usepackage{ae,aecompl}
\usepackage{mathptmx,pdflscape,multicol}
\usepackage{txfonts,nameref}
\definecolor{LightCyan}{rgb}{0.88,1,1}
\definecolor{skyblue}{RGB}{0, 170, 255}
\definecolor{BlueViolet}{RGB}{138, 43,226}
\definecolor{MidnightBlue}{RGB}{0, 120, 255}
\definecolor{orange}{RGB}{255,140,0}
\definecolor{forestgreen}{RGB}{20,160,120}
\definecolor{gray50}{RGB}{127,127,127}
\def\HI{{\ion{H}{I}}}
\def\HII{{\ion{H}{II}}}
\def\xb{\bar{x}_{\rm \ion{H}{I}}}
\def\Tb{{T_{\rm b}}}
\def\tTb{\tilde{T}_{\rm b}}
\def\tTt{\tilde{T}_{\rm t}}
\def\tTn{\tilde{T}_{\rm N}}

\def\kk{\mathbfit{k}}
\def\xx{\mathbfit{x}}
\def\cov{\mathbfss{C}}

\def\signal{EoR 21-cm signal~}

\def\tri{\bar{T}(k_i,k_j)}
\def\try{T(\kk_{g_{i}},-\kk_{g_{i}},\kk_{g_{j}},-\kk_{g_{j}})}

\def\mpc{~\rm Mpc}
\def\impc{~\rm Mpc^{-1}}
\def\Mmin{M_{\rm min}}
\def\Nion{N_{\rm ion}}
\def\Rmfp{R_{\rm mfp}}
\def\Nk{N_{\rm k_{i}}}

\title[EoR 21-cm power spectrum error covariance]{The impact of non-Gaussianity on the error covariance for observations of the  Epoch of Reionization 21-cm power spectrum}

\author[A. K. Shaw, S. Bharadwaj and R. Mondal]{
	Abinash Kumar Shaw$^{1,2}$\thanks{E-mail:\href{mailto:abinashkumarshaw@iitkgp.ac.in}{abinashkumarshaw@iitkgp.ac.in}},
	Somnath Bharadwaj$^{1,2}$ and
	Rajesh Mondal$^{3}$
	\\
	$^{1}$Department of Physics, Indian Institute of Technology Kharagpur, Kharagpur 721302, India\\
	$^{2}$Centre for Theoretical Studies, Indian Institute of Technology Kharagpur, Kharagpur 721302, India\\
	$^{3}$Astronomy Centre, Department of Physics and Astronomy, University of Sussex, Brighton BN1 9QH, UK}

\date{Accepted: 2019 May 31; Revised: 2019 May 27; Received: 2019 February 22}
\pubyear{2019}

\begin{document}
\label{firstpage}
\pagerange{\pageref{firstpage}--\pageref{lastpage}}
\maketitle

\begin{abstract}
Recent simulations show the Epoch of Reionization (EoR) 21-cm signal to be inherently non-Gaussian whereby the 
error covariance matrix $\cov_{ij}$ of the 21-cm power spectrum (PS) contains a trispectrum contribution that would be absent if the signal were  Gaussian.
 Using  the binned power spectrum and trispectrum  from simulations, 
here we present a methodology for incorporating  these with the  baseline distribution and system noise to make error predictions for observations with any  radio-interferometric array. Here we consider the upcoming SKA-Low. Non-Gaussianity  enhances the errors introducing a positive deviation $\Delta$ relative to the Gaussian predictions. $\Delta$ increases with observation time $t_{\rm obs}$ and saturates as the errors approach the cosmic variance. Considering  $t_{\rm obs}=1024$ hours  where a $5 \sigma$ detection is possible at all redshifts $7 \le z \le 13$, in the absence of foregrounds we find that the deviations are important at small $k$ where  we have $\Delta \sim 40-100 \%$ at $k  \sim 0.04 \impc$ for some of the redshifts and also at intermediate $k \, (\sim 0.4 \impc)$ where we have $\Delta \sim 200 \%$ at  $z=7$. Non-Gaussianity also introduces correlations between the errors in different $k$ bins, and we find both correlations and anticorrelations with the correlation coefficient value spanning  $-0.4 \le r_{ij} \le 0.8$. Incorporating the foreground wedge,  
 $\Delta$ continues to be important ($> 50 \%$) at $z=7$.  We conclude that non-Gaussianity makes a significant contribution to the errors and this is important in the context of the future instruments that aim to achieve high-sensitivity measurements of the EoR 21-cm PS.  
\end{abstract}

\begin{keywords}
cosmology: reionization, first stars, large-scale structure of universe, diffuse radiation, methods: statistical, technique: interferometric
\end{keywords}


\section{Introduction}
\label{sec:intro}
The Epoch of Reionization (EoR) is an important but poorly understood milestone in the cosmic history when the hydrogen in the universe underwent a transition from neutral (\HI) to ionized (\HII) phase. Our current knowledge of the EoR comes from several indirect observations. The measurements of the Thomson scattering optical depth $\tau_{\rm Th}=0.058\pm0.012$ \citep{Planck_2016, 2016_planck} of the cosmic microwave background radiation (CMBR) with the free electrons in the intergalactic medium (IGM) suggests that the universe was ionized at less than $10\%$ level at redshifts above $z\sim 10$. Measurements of the high-redshift quasar spectra \citep{becker_2001, Fan_2002, Fan_2006, gallerani_2006, Becker_2015} show a complete Gunn--Peterson trough and also measurements of the Gunn--Peterson optical depth $\tau_{\rm GP}$ suggest that the reionization was over by $z\sim 6$. Recent studies of the Ly-$\alpha$ emitters (LAE) show a rapid decline in the luminosity function at $z\geq 6$ \citep{Ouchi_2010, Jensen_2014, Konno_2014, Faisst_2014, santos_Lya-2016, Ota_2017, Zheng_2017} which suggests a rapid increase in the \HI~ density in the IGM and a patchy \HI~ distribution at those redshifts. These indirect observations together suggest the reionization to occur within a redshift range $6\leq z \leq 12$ \citep{mitra_2013,Robertson_2013,mitra_2015,Robertson_2015, Mondal_I,Dai_2018}. However such indirect observations are not adequate to address many fundamental issues related to the EoR such as the exact duration and timing, the properties of the ionizing sources and the topology of \HI~ distribution.

Observations of the redshifted 21-cm radiation due to the hyperfine transition of \HI~is a promising probe to study the high-redshift universe \citep{sunyaev_1972, scott_1990}. The low-frequency radio interferometers will measure brightness temperature fluctuations of the EoR 21-cm radiation \citep{Bharadwaj2001,Bharadwaj_2005}. A substantial effort is currently underway to measure the \signal using the first-generation radio interferometers e.g. GMRT\footnote{\url{http://www.gmrt.ncra.tifr.res.in}} \citep{Paciga_2013}, MWA\footnote{\url{http://www.haystack.mit.edu/ast/arrays/mwa}} \citep{Jacobs_2016}, LOFAR\footnote{\url{http://www.lofar.org}} \citep{Yatawatta_2013}, PAPER\footnote{\url{http://eor.berkeley.edu}} \citep{PAPER_2014} and the second-generation interferometers such as HERA\footnote{\url{http://reionization.org}} \citep{Pober_2014,Ewall-Wice} and the upcoming gigantic SKA\footnote{\url{http://www.skatelescope.org}} \citep{koopmans2014}. These experiments aim to measure the EoR 21-cm power spectrum (PS) \citep{Ali_2004}. The expected \signal is about $4-5$ orders of magnitude weaker compared to the galactic and extragalactic foregrounds \citep{Ali_2008, Bernardi_2009, Bernardi_2010, Abhik_2012, Paciga_2013, Beardsley_2016}. The foregrounds, together with the system noise and other calibration errors, pose a huge challenge for the measurement of the EoR 21-cm PS. Only weak upper limits on the EoR 21-cm PS have been estimated till date \citep{McGreer_2011,Parsons_2014,Pober_2016}. In addition to the PS, various other statistics such as the variance \citep{patil_2014}, bispectrum \citep{Yoshiura_2015,bispec_shimabukuro,Majumdar_2018} and the Minkowski Functional \citep{Akanksha_2017,Bag_2018} have been proposed to quantify the \signal. 

In the recent past, several works have made quantitative predictions of the sensitivity for measuring the EoR 21-cm PS \citep{Morales_2004}. \citet{McQuinn_2006} have made predictions for $1000$ hours of observations with the MWA, LOFAR and the upcoming SKA-Low. \citet{Beardsley_2013} have estimated that MWA is capable of detecting the \signal at $\sim 14\sigma$ level with $\sim 900$ hours of observations. \citet{Zaroubi_2012} have made quantitative predictions for sensitivity of LOFAR considering $600$ hours of observations, and \citet{Jensen_LOFAR} have predicted that LOFAR will be able to detect the EoR 21-cm PS at $k \sim 0.1 \impc$ with $\sim 1000$ hours of observations.  \citet{Parsons_2012b} have predicted  that the \signal can be detected at  $k \sim 0.2h  \impc$ with PAPER in $7$ months of observations. The results of \citet{Pober_2014} suggest that the upcoming HERA will be able to detect the EoR 21-cm PS at a level $\sim 30\sigma$ within the $k$ range $0.1-1 \impc$ assuming a moderate foreground model. \citet{Ewall-Wice} have studied the prospects of detecting the EoR 21-cm PS with HERA incorporating X-ray heating of the IGM.

The upcoming SKA-Low, to be located in Australia, will be the most sensitive radio telescope to be built. It will have $512$ stations, each of which combines the signal from several constituent log periodic dipole antennas. Each of these station is planned to be  $\sim 35 ~{\rm m}$ in diameter. The telescope will operate within a frequency band of $50-350~\rm{MHz}$ and it will have $\sim 20~\rm{deg^2}$ field of view. The interferometer will have a compact core and $3$ spiral arms which will extend up to a large distance such that maximum antenna separation is $\sim 60~{\rm km}$. A recent study by \citet{Mellema_2013} has quantified the prospects of detecting the EoR 21-cm PS with SKA-Low. The authors have predicted the errors in the measured EoR 21-cm PS at three different redshifts $8,~10,~\text{and}~12$. In this analysis they have  varied the number of  core antennas and also the core radius. The analysis incorporates the system noise assuming $1000$ hours of observation with a bandwidth of $10~{\rm MHz}$.  They find that it will be possible to achieve a maximum SNR of $\sim 100$ at $k\sim 0.4 \impc$ for all the three redshifts. They also find that the predictions for SKA-Low show a significant improvement in comparison with other precursor telescopes such as MWA, LOFAR and PAPER \citep[figs 21 and 22 of][]{Mellema_2013}.

All the existing  predictions for detecting the EoR 21-cm PS have assumed the signal to be a Gaussian random field. This assumption plays a crucial role in making the predictions. The PS completely specifies the statistical properties of the signal for a Gaussian random field, and this assumption 
allows the signal in each Fourier mode to be treated as being independent. Gaussianity is possibly a good assumption during the early stages of EoR, and also when one observes very large length-scales. However, the growth and subsequent overlapping of the \HII~ regions make the signal highly non-Gaussian as reionization progresses \citep{Pandey_2005}. The PS no longer quantifies the entire statistical properties of the signal as the signal in different Fourier modes are correlated. Higher order statistics like the bispectrum \citep{Majumdar_2018} and trispectrum are needed to quantify these correlations.  This also affects the error predictions for the PS. Considering only cosmic variance (CV) that is inherent to the signal, \citet{Mondal_2015} have studied  the effects of  non-Gaussianity on the error predictions for the EoR 21-cm PS. For a Gaussian random field, the SNR for the 21-cm PS is expected to increase as the square root of the number of independent Fourier modes. However, \citet{Mondal_2015} find that as a consequence of the non-Gaussianity  the SNR saturates at a limiting value [SNR]$_{l}$ beyond which it does not increase any further. The value of [SNR]$_{l}$ was also found to decreases with the progress of reionization that corresponds to an increase in the non-Gaussianity. Two subsequent papers \citep{Mondal_I,Mondal_II} have quantified the error covariance for the binned PS, which now has an extra contribution from the trispectrum as compared to the Gaussian situation where the error covariance can be expressed entirely in terms of the PS. In these papers they have developed a unique statistical technique for estimating the bin-averaged trispectrum from the PS error covariance. They have used an ensemble of seminumerical EoR simulations to estimate the error covariance and the trispectrum at several redshifts in the range $7 \le z \le 13$. The trispectrum contribution is found to increase significantly as reionization progresses. The non-Gaussianity is found to result in larger error estimates compared to the Gaussian predictions. Non-Gaussianity also introduces correlations between the PS error estimates at different bins.

In this paper, we predict the prospects of measuring the EoR 21-cm PS using observations with the upcoming SKA-Low. To this end we study the error covariance of the EoR 21-cm PS that will be measured by SKA-Low. Unlike the previous works (e.g. \citealt{Mellema_2013}), our analysis incorporates the inherent non-Gaussian nature of the signal. We have used the EoR 21-cm PS and trispectrum from the simulations of \citet{Mondal_II}. We include the system noise contribution to calculate the full PS error covariance for the current proposed configuration of SKA-Low\footnote{\label{ft:ska}\href{https://astronomers.skatelescope.org/wp-content/uploads/2016/09/SKA-TEL-SKO-0000422_02_SKA1_LowConfigurationCoordinates-1.pdf}{SKA1\_LowConfigurationCoordinates-1.pdf}}. The analysis in this paper also incorporates the impact of foregrounds considering the \signal to be free of other possible calibration errors. 

The structure of this paper is as follows. Section \ref{sec:sim} briefly describes the simulations and the techniques used in \citet{Mondal_II} to obtain the EoR 21-cm PS and trispectrum. Section \ref{sec:error_cov} briefly presents the SKA-Low configuration and discusses how to combine the observed visibility data for an optimal estimate of the EoR 21-cm PS. We also present a framework to compute the EoR 21-cm PS error covariance. Section \ref{sec:res} presents the results considering no foregrounds. In Section \ref{sec:fgd} we study the effects of foregrounds and finally summarize and discuss our findings in Section \ref{sec:dis}. In keeping with the simulations of \citet{Mondal_II}, we have used the Planck+WP \citep{planck_2014} best-fitting cosmological parameters throughout this paper.


 \section{Simulating The E\lowercase{o}R 21-\lowercase{cm} Signal}\label{sec:sim}
 We have simulated the \signal at six different redshifts $z=13,11,10,9,8~\text{and}~7$ using a seminumerical technique (\citealt{Majumdar_2013,Mondal_2015}) that comprises three major steps. First, we generate the dark matter distributions at the aforementioned redshifts using a publicly available particle mesh \textit{N}-body code\footnote{\url{https://github.com/rajeshmondal18/N-body}} \citep{Srikant2004}. We have simulated the dark matter distributions within a cube of comoving volume $V=[215.04\mpc]^3$ with a grid size of $0.07\mpc$ and a mass resolution of $1.09\times 10^8~M_\odot$. Next, we identify the dark matter halos within the matter distribution using a publicly available halo finder\footnote{\url{https://github.com/rajeshmondal18/FoF-Halo-finder}} based on the Friends-of-Friend (FoF) algorithm \citep{davis} with a linking length $0.2$ times the mean inter-particle spacing and a minimum halo mass of $1.09\times 10^9~M_\odot$ which corresponds to $10$ simulation particles. In the final step we generate the reionization map using a publicly available seminumerical code\footnote{\url{https://github.com/rajeshmondal18/ReionYuga}} following the formalism adopted by \citet{Choudhury_2009}. We assume that the hydrogen traces the dark matter, and the haloes with masses exceeding a minimum halo mass $\Mmin$  ($M\geq\Mmin$) host the ionizing sources, the number of ionizing photons $N_\gamma$ emitted by a source being   proportional to the  host halo mass $M$ through a dimensionless  constant of proportionality  $\Nion$, which incorporates a large number of unknown parameters like the star formation efficiency and the UV photon escape fraction. 
 
 The hydrogen and photon densities are, respectively, smoothed over spheres of radius $R$. Any grid point within the simulation is considered to be completely ionized if the smoothed photon density exceeds the smoothed hydrogen density, the smoothing radius is allowed to vary from one grid spacing to a maximum value of $\Rmfp$.   The resulting \HI~ distribution is mapped to redshift space using the prescription of \citet{Majumdar_2013} to generate the final 21-cm  brightness temperature distribution on a grid eight times coarser than the \textit{N}-body simulation. The simulations used here are exactly the same as those that were used in \citet{Mondal_I,Mondal_II} and the reader is referred to there for further details. There simulations have three free parameters namely $\Mmin$ the minimum halo mass, $\Nion$ the ionizing efficiency and $\Rmfp$ the mean free path of the ionizing photons. We have used the values $\Mmin = 1.09\times 10^9~M_\odot$, $\Nion = 23.21$ and $\Rmfp = 20\mpc$ \citep{cowie} to obtain a reionization history where the mean mass averaged neutral fraction has a value $\xb = 0.5$ at $z=8$ and is over by $z\sim 6$. The integrated Thomson scattering optical depth obtained using these parameter values, $\tau=0.057$, is also consistent with the observations \citep{Planck_2016} where $\tau=0.058\pm 0.012$.

 
 \section{Power Spectrum Error Covariance}\label{sec:error_cov}
  We quantify the statistics of the EoR  21-cm brightness temperature fluctuations using the power spectrum (PS) which is defined as $P(k)=V^{-1} \langle \tTb(\kk) \tTb(-\kk) \rangle$. Here $V$ is the simulation (observational) volume, $\tTb(\kk)$ is the Fourier transform of the brightness temperature fluctuations $\delta\Tb(\xx)$ and $\kk$ is a wave vector. In the absence of foregrounds and calibration errors, the  brightness temperature fluctuations recorded by a radio interferometer is  $\tTt(\kk)=\tTb(\kk)+\tTn(\kk)$  which is a sum of the 21-cm signal $\tTb(\kk)$ and  the system noise contribution $\tTn(\kk)$ . The PS corresponding to $\tTt(\kk)$ therefore is a sum of $P(k)$ and $P_{\rm N}(k)$ which is the system noise PS {\it i.e.}  $P_{\rm t}(k)= [P(k)+P_{\rm N}(k)]$. We have used the simulations described in Section \ref{sec:sim} to predict the EoR 21-cm PS $P(k)$.  In this work we make predictions for the upcoming SKA-Low\textsuperscript{\ref{ft:ska}}, and we have used the specification described in the subsequent paragraph to compute the noise PS $P_{\rm N}(k)$. We have considered the upcoming SKA-Low to be an array of $512$ stations\textsuperscript{\ref{ft:ska}}, each of which is a station of diameter $D=35$ m. The instrument will operate within a frequency range of $50-350~{\rm MHz}$ which will probe the \HI~ 21-cm signal between $z=27$ and $z=3$.  The \signal evolves significantly along the line of sight (LoS) and observations at  different  redshifts will probe the signal at different stages of reionization due to the light-cone effect \citep{Light_cone_I,Light_Cone_II}. As a consequence, the signal no longer remains ergodic along the LoS and there is a significant loss of information if the entire frequency band is used to estimate the PS \citep{Rajesh_Light-cone,Mondal_2019}. In the present work we have avoided this by restricting the analysis to six different redshift slices each of width $\Delta z=0.75$ centred at redshifts $z=13,~11,~10,~9,~8~\text{and}~7$.  We have also assumed that the entire frequency  bandwidth is divided into frequency channels of width $\Delta \nu_c = 0.1~{\rm MHz}$. Note that the antenna layout, the number of antennas and the channel width $\Delta \nu_c$ assumed here are only representative values, and may change in the final implementation of the telescope.

The analysis in this paper considers an observation tracking a field at declination DEC$=-30^\circ$ using SKA-Low for $8$ hours with $60$-second integration time. The $60$-second integration time has been chosen here to keep the simulated baseline data volume small. However, the purpose of simulating the array baseline configuration here is to primarily estimate  $P_{\rm N}(k)$, and we find that the noise predictions do not show any noticeable change even when the integration time is reduced to $30$ seconds or to $15$ seconds. Considering $\mathbfit{d}$ to be the projection of the antenna separation on the plane perpendicular to the LoS, we use $\mathbfit{U}= \mathbfit{d}/\lambda_c$ with $\lambda_c$ being the wavelength that corresponds to the central frequency $\nu_c$ of a slice. The subsequent analysis is restricted to the baselines $\mathbfit{U}$ corresponding to the antenna separations $|\mathbfit{d}| \leq 19~{\rm km}$ as the baseline distribution falls off rapidly at larger values of $\mathbfit{d}$. The simulated observations provide us the baselines $\mathbfit{U}_i$ and frequency channels $\nu_n$ at which the signal will be measured. We use $\kk_{\perp i} = (2\pi \mathbfit{U}_i)/r_c$ and $k_{\parallel m}=(2 \pi m)/(r_c^\prime B)$ with $0 \leq m \leq N_c/2$ where $r_c$ is the comoving distance to the centre of a redshift slice, $r_c^\prime = \partial r/ \partial \nu \big|_{\nu=\nu_{c}}$, $B$ is the frequency bandwidth of the  redshift slice  and $N_c= B/\Delta \nu_c$. Note that $k_{\parallel m}$ is the Fourier conjugate of $r_{c}^{\prime}(\nu_n-\nu_c)$. The simulations provide us with a set of comoving vectors $(\kk_{\perp i},~k_{\parallel m})$  at which we will obtain measurements of the brightness temperature fluctuations $\tTb(\kk_{\perp i},~k_{\parallel m})$. Two different baselines having separation less than $D/\lambda_c$ do not have independent information due to overlap of the antenna beam pattern \citep{Bharadwaj_2005}. We grid the comoving wave vectors  with a grid of size $\Delta k_x=\Delta k_y= (2 \pi D)/(\lambda_c r_c)$ and $\Delta k_z= (2 \pi)/(r_c^\prime B)$. Considering a grid point $\kk_g$, we define $\tau(\kk_g)$ to be the number of measurements that lie within a voxel centred at $\kk_g$. We use $\tau(\kk_g)$ to estimate the noise PS $P_{N}(\kk_g)$ at each grid point $\kk_g$ using the following expression \citep{SumanCh_2019}:
\begin{equation}\label{eq:Pn}
P_{\rm N}(\kk_g) = \frac{r_c^2 ~r_c^\prime ~T_{\rm sys}^2 ~\lambda_c^2}
{N_p~ N_t~ \Delta t ~\chi~ A_g ~\tau(\kk_g)} = \frac{8 ~{\rm hours}}{t_{\rm obs}}\times \frac{P_0}{\tau(\kk_g)}~.
\end{equation}
Here $T_{\rm sys}$ is the system temperature, $N_p$ is the number of polarizations, $N_t$ is the number of observed nights with $8$ hours per night, $\Delta t$ is the integration time, $A_g=(\pi D^2)/4$ is the geometric area of a single antenna. It is convenient to quantify the  total duration of the observations using $t_{\rm obs}=N_t \times 8$ hours instead of $N_t$, and we have used $t_{\rm obs}$ through the subsequent discussion of this paper. The system temperature $T_{\rm sys}= T_{\rm sky} +T_{\rm rec}$ is a sum of the sky temperature $T_{\rm sky}= 60 \lambda^{2.55}~{\rm K}$ \citep{Fixsen_2011} and the receiver temperature $T_{\rm rec}=100 ~{\rm K}$. Here $\chi$ is defined using 
\begin{equation}
    \frac{1}{\chi}=\frac{A_g}{\lambda_c^2} \frac{[\int d \Omega \, A(\bm{\theta})]^2}{[\int d \Omega 
    \, A^2(\bm{\theta})]}
    \label{eq:c1it}
\end{equation}
where $A(\bm{\theta})$ is the telescope's primary beam pattern \citep{Tapomoy_2013,Parsons_2014}. We have approximated the beam pattern with a Gaussian  $e^{-(\theta/\theta_0)^2}$ \citep{Samir_2014} and evaluated the solid angle integral in the flat sky approximation to obtain $\chi=0.53$. Note that $P_{\rm N}(\kk_g)$ is infinitely large at the grid points where $\tau(\kk_g)=0$ \textit{i.e.} the grid points that are not sampled by the telescope baseline distribution.

Considering a typical SKA-Low observation spanning an angular extent of $\sim 3^\circ\times 3^\circ$ on the sky with an angular resolution $\sim 1^\prime$ and a frequency bandwidth of $\sim 64~ \rm MHz$ with frequency resolution $\sim 0.1 ~ \rm MHz$, this corresponds to $N_{k}=[180\times 180 \times 640] \simeq 2\times 10^7$ different grid points at which the EoR 21-cm PS will be  measured. The dimension of the resulting PS error covariance matrix is  $\sim 10^7 \times 10^7$ which renders further computations prohibitively expensive if not impossible. In order to overcome the intractability of such a large covariance matrix, we bin the $\kk$ space and use the binned PS estimator that, for the $i$-th bin, is defined as 
\begin{equation}
    \hat{P}_{\rm t}(k_i)={V}^{-1}~ \sum_{g} w_g \tTt(\kk_g) \tTt(-\kk_g)~,
 \label{eq:x1}
 \end{equation}  
where the sum is over the $\kk_g$ modes within the $i$-th bin and $w_g$ is the normalized weight associated with each mode with $\sum_{g} w_g =1$. Here $k_i=\sum_g  w_g k_g$ is the average $k$ value corresponding to the $i$-th bin. The weights $w_g$ have been introduced to account for the fact that the ratio $P(\kk_g)/P_N(\kk_g)$ varies across the different grid points, and as discussed later, the weights have been chosen so as to maximize the SNR of the bin-averaged PS. For the present analysis we have divided the available $\kk$ space into $10$ logarithmic spherical bins. The ensemble average of $\hat{P}_{\rm t}(k_i)$ gives the bin-averaged PS $\bar{P}_{\rm t}(k_i) =\langle \hat{P}_{\rm t}(k_i) \rangle=\bar{P}(k_i)+\bar{P}_{\rm N}(k_i)$. Note that the resulting estimate has a noise bias $\bar{P}_{\rm N}(k_i)$, this however can be eliminated by suitably modifying the estimator \citep{TGE}. In the subsequent analysis we assume that the noise bias has been eliminated and we have an unbiased estimate of the bin-averaged power spectrum $\bar{P}(k_i)$. The noise contribution to the PS error covariance $\cov_{ij} =\langle [\hat{P}_{\rm t}(k_i)-\bar{P}_{\rm t}(k_i)] [\hat{P}_{\rm t}(k_j)-
\bar{P}_{\rm t}(k_j)] \, \rangle$, however, cannot be eliminated and following the calculation presented in \citet{Mondal_I}, we have 
\begin{equation}
\begin{split}
 \cov_{ij}= 
 \sum_{g_{i}}^{} w_{g_{i}}^2 &[P(\kk_{g_{i}})+P_{\rm N}(\kk_{g_{i}})]^2~ \delta_{ij} \\ &+{V}^{-1} \sum_{g_{i}}^{} \sum_{g_{j}}^{} w_{g_{i}} w_{g_{j}} T(\kk_{g_{i}}, -\kk_{g_{i}}, \kk_{g_{j}},-\kk_{g_{j}})~,
\end{split}
\label{eq:cov}
\end{equation}
where the sum is over the grids points  $\kk_{g_{i}}$ and $\kk_{g_{j}}$ in the $i$-th and the $j$-th bins respectively. The trispectrum $T(\kk_1, -\kk_2, \kk_3,-\kk_4)$ originates  due to non-Gaussianity of the EoR 21-cm signal, the quantity that appears here is the weighted bin-averaged trispectrum. For the diagonal terms of the covariance matrix $\cov_{ij}$ the trispectrum quantifies the excess with respect to the Gaussian predictions. The off-diagonal terms of  $\cov_{ij}$ are predicted to be zero if the EoR 21-cm signal were a Gaussian random field. The trispectrum arising due to the non-Gaussianity introduce non-zero off-diagonal terms corresponding to correlations (and anticorrelations) between the errors in the PS estimates in the different $k$ bins \citep{Mondal_I,Mondal_II}. The system noise has been considered to be outcome of a Gaussian random process and this does not contribute to the non-Gaussianity through the trispectrum.

\subsection{Computing the Error Covariance from the Simulations}
 The PS error covariance $\cov_{ij}$ consists of two components : (1) the cosmic variance (CV), and (2) the system noise. According to equation (\ref{eq:cov}), we need the EoR 21-cm PS $P(\kk_g)$, the EoR 21-cm trispectrum $\try$, the noise PS $P_{\rm N}(\kk_g)$ and appropriate weights $w_g$ to compute the $\cov_{ij}$. The reionization simulations of \citet{Mondal_II} provide us the bin-averaged EoR 21-cm PS  
 \begin{equation}
 \bar{P}(k_i)=\Nk^{-1} \,   \sum_{g_{i}} P(\kk_{g_{i}})  
 \label{eq:d1}
 \end{equation}
  and the bin-averaged trispectrum
 \begin{equation}
  \bar{T}(k_i,k_j)= (N_{\rm k_i} \, N_{\rm k_j})^{-1}  \sum_{g_{i}}^{} \sum_{g_{j}}^{} \try ~,
 \label{eq:d2}
 \end{equation}
 where the sum in equation~(\ref{eq:d2}) is over the grid points ($\kk_g$ modes) in the $i$-th and $j$-th bins, and the $N_{\rm k_i}$ and $N_{\rm k_j}$ are numbers of grid points in the respective bins. The bins that we have chosen to analyse  the simulated SKA-Low observations have exactly the same boundaries as the bins used to analyse the EoR simulations in \citet{Mondal_II}, however we cannot directly use the $\bar{P}(k_i)$ and $\bar{T}(k_i,k_j)$ from \citet{Mondal_II} in equations (\ref{eq:x1}) and (\ref{eq:cov}) to predict the PS error covariance for the SKA-Low observations. First, equations (\ref{eq:d1}) and (\ref{eq:d2}) assume uniform weights, whereas it is necessary to consider the variation of $w_g$ across the grid points to account for the non-uniform sampling when considering the simulated observations (equations~\ref{eq:x1} and \ref{eq:cov}). Further, the resolution of the simulations and the observations will, in general, be different and consequently the $\kk$ grid spacing will also differ.

One can attempt to  estimate  the ensemble averages of $P(\kk_g)$ at every  individual grid point and $\try$ at every pair of grid points, however these estimates will be extremely noisy due to the limited number of statistically independent realizations in the \signal ensemble (e.g. $50$ in \citealt{Mondal_II}). Further, we have an enormous volume of the trispectrum data that renders this approach unfeasible. The issue now is to predict the bin-averaged PS (equation~\ref{eq:x1}) and its error covariance (equation~\ref{eq:cov}) for the SKA-Low observations using the results (equations~\ref{eq:d1} and \ref{eq:d2}) from the simulations of \citet{Mondal_II}.

Here we have assumed that the EoR  21-cm PS does not vary much across the grid points $\kk_{g_{i}}$ within a bin (say the $i$-th bin), and in equations (\ref{eq:x1}) and (\ref{eq:cov}) we have used the simulated $\bar{P}(k_i)$ from \citet{Mondal_II} to calculate $P(\kk_{g_{i}})= \bar{P}(k_i)$ for all the grid 
points in the  $i$-th bin. The value of $\try$ in equation  (\ref{eq:cov})  depends on the magnitude and direction of the two vectors $\kk_{g_{i}}$ and $\kk_{g_{j}}$, and both of these can vary widely even when the two vectors are in the same bin ($i=j$). An even wider variation is possible when the two vectors are in two different bins $i$ and $j$. Unfortunately this information is not available in $\bar{T}(k_i,k_j)$ (equation~\ref{eq:d2}) evaluated from the simulation of \citet{Mondal_II}. Here we have considered  two different assumptions regarding the trispectrum at two different modes $\kk_{1}$ and $\kk_{2}$. These two assumptions correspond to two extreme cases. Case--I: we assume that  all the modes within a bin are equally correlated {\em i.e.} $T(\kk_1,-\kk_1,\kk_2,-\kk_2)= {T}_c(k_i,k_i)$  when both $\kk_1$ and $\kk_2$ are in the $i$-th bin, and the correlation between modes in two different bins does not depend on the magnitude or orientation of the individual vectors {\em i.e.} $T(\kk_1,-\kk_1,\kk_2,-\kk_2)= {T}_c(k_i,k_j)$  when $\kk_1$ and $\kk_2$ are in the $i$-th and $j$-th bins, respectively. Case--II: we assume that the signal in two different Fourier modes is uncorrelated unless $\kk_1=\kk_2$ {\em i.e.} $T(\kk_1,-\kk_1,\kk_2,-\kk_2)= \delta_{\kk_1,\,\kk_2} {T}_u(k_i,k_i) $ when the mode $\kk_i$ is in the $i$-th bin. Case--I corresponds to the situation in which we have the maximum possible correlation between different modes whereas Case--II corresponds to the situation in which we have the minimum possible correlation between two different modes. In reality we expect the correlation between two modes to vary with the separation between the two modes, and the result is expected to lie within the two extreme  cases considered here. Considering  equation~(\ref{eq:d2}),  we obtain  $T_{\rm c}(k_i,k_j)=\tri$ for Case--I whereas it predicts $T_{\rm u}(k_i) = \Nk \bar{T}(k_i,k_i)$ for Case--II. Note that Case--II predicts the error covariance to be completely diagonal with all the off-diagonal terms being zero  which is inconsistent with the findings of \citet{Mondal_I}. While Case--II is unrealistic for the off-diagonal elements of the covariance matrix, we still consider its predictions for the diagonal elements in order to illustrate  the effect of  partial decorrelation in the value of the trispectrum across different modes.

We calculate the weights separately for both the cases by extremizing the SNR$= \bar{P}(k_i)/\sqrt{\cov_{ii}}$ with respect to $w_g$. Considering Case--I the unnormalized weights that extremizes the SNR are
\begin{equation}\label{eq:wt1}
\tilde{w}_{g_{i}}= \frac{1}{[\bar{P}(k_i)+P_{\rm N}(\kk_{g_{i}})]^2}~,
\end{equation}
which have $P_{\rm N}(\kk_{g})$ in the denominator, \textit{i.e.} the grid points with higher noise contribute less to the bin averaged quantities. The grid points $\kk_{g}$, which are unsampled during observations, \textit{i.e.} $\tau(\kk_{g})=0$, have $P_{\rm N}(\kk_{g})=\infty$ (equation \ref{eq:Pn}). The weight $\tilde{w}_{g}=0$ (equation \ref{eq:wt1}) for the unsampled  grid points  and they do not contribute to the bin averaged quantities. Using equation~(\ref{eq:wt1}) in equation~(\ref{eq:cov}), we obtain  the corresponding PS error covariance matrix 
\begin{equation}
\cov_{ij}= \frac{1}{\sum_{g_{i}}^{} \tilde{w}_{g_{i}}} \delta_{ij} + \frac{\tri}{V}~.
\label{eq:d3}
\end{equation}
For comparison we consider  the error covariance for a situation where the signal is a Gaussian random field for which the trispectrum is zero. The weights $\tilde{w}_{g_{i}}$ here are unchanged and these are given by equation (\ref{eq:wt1}), and we have the PS error covariance matrix  
\begin{equation}
\cov_{ij}^{\rm G}= \frac{1}{\sum_{g_{i}}^{} \tilde{w}_{g_{i}}} \delta_{ij}~.
\label{eq:dg3}
\end{equation}
The diagonal terms of the covariance matrices (equations \ref{eq:d3} and \ref{eq:dg3}) predict the error variance in the measured EoR 21-cm PS, \textit{i.e.} $\cov_{ii}=\langle [\Delta \hat{P}(k_i)]^2 \rangle$. Equations (\ref{eq:d3}) and (\ref{eq:dg3}) indicate that the Gaussian consideration underestimates the variance of the measured PS. The off-diagonal terms of the covariance matrix ($i \neq j$) predict the correlation between the errors at the $i$-th and $j$-th bins $\cov_{ij}=\langle [\Delta \hat{P}(k_i) \Delta \hat{P}(k_j)]\rangle$. The off-diagonal terms are zero for a Gaussian random field, and the errors in the different bins are uncorrelated. Non-Gaussianity however may introduce correlations between the different bins through the off-diagonal components of the trispectrum.

We first discuss the diagonal terms $\cov_{ii}$, \textit{i.e.} the variance. This  has contributions from the CV as well as the system noise. The noise PS $P_{\rm N}(\kk_{g_{i}})$ scales as $t_{\rm obs}^{-1}$ (equation \ref{eq:Pn}) and this has a large value for small observation times. Considering the  behaviour of $\cov_{ii}$,  for small observation times this is governed by the system noise contribution and we have 
\begin{equation}
   \cov_{ii}\simeq \left(\frac{8~{\rm hours}}{t_{\rm obs}}\right)^2 \times \frac{P_0^2}{\sum_{g_{i}}[\tau(\kk_{g_{i}})]^2}~.
   \label{eq:e1}
\end{equation}
Equation (\ref{eq:e1}) shows that $\cov_{ii}\propto t_{\rm obs}^{-2}$ and consequently SNR$\propto t_{\rm obs}$ for small observation times. The observations with very large $t_{\rm obs}$ elucidate another extreme of the error estimates (equation \ref{eq:d3}) where $P_{\rm N}(\kk_{g})\simeq 0$,  and $\cov_{ii}$ converges to the `CV' that is given by 
\begin{equation}
    \cov_{ii}=\frac{\bar{P}^2(k_i)}{N_{g_{i}}} + \frac{\bar{T}(k_i,k_i)}{V}~.
    \label{eq:e4}
\end{equation}
where $N_{g_{i}}$ is the number of sampled grid points in the $i$-th bin. The CV represents the lower limit for the PS error variance. This arises due to the inherent statistical uncertainty in the EoR 21-cm signal. The actual predicted error variance  for a finite observing time will typically be larger than this due to the system noise contribution.

The corresponding cosmic variance  for a Gaussian random field  (equation \ref{eq:dg3}) is given by 
\begin{equation}
    \cov_{ii}=\frac{\bar{P}^2(k_i)}{N_{g_{i}}}~.
    \label{eq:eg4}
\end{equation}
A comparison of equations (\ref{eq:e4}) and (\ref{eq:eg4}) illustrates an important difference between the Gaussian and non-Gaussian situations. We see that it is possible to reduce the CV with no lower bound by combining the signal from a larger number of $\kk$ modes in the bin, \textit{i.e.} increasing $N_{g_{i}}$. In contrast, the presence of the trispectrum in equation (\ref{eq:e4}) sets a lower limit to the value of $\cov_{ii}$, and it is not possible to lower the variance any further by increasing the number of $\kk$ modes
\citep{Mondal_2015}.

Next considering the off-diagonal terms $\cov_{ij}=\tri/V$ (equation \ref{eq:d3})  which quantify the correlation  between different bins, we see that this only depends on the trispectrum. This is intrinsic to the signal, and therefore is independent of the system noise and observation time. 

 Considering Case--II, the unnormalized weights are given by 
 \begin{equation}\label{eq:wt2}
 \tilde{w}_{g_{i}}= \frac{1}{[\bar{P}(k_i)+P_{\rm N}(\kk_{g_{i}})]^2+ {\Nk {V}^{-1}\, 
 \bar{T}(k_i,k_i)}}~,
 \end{equation} 
 which differ from the weight  in Case--I (equation \ref{eq:wt1}). The weights now include a contribution  from the trispectra for the non-Gaussian signal. Here also the weights are zero for the grid points that are not sampled by the baseline distribution. The weights for  Case--II  match those for  Case--I (equation \ref{eq:wt1}) if the signal were a Gaussian random field. The PS error covariance (using equations \ref{eq:cov} and \ref{eq:wt2}) in Case--II is   given by 
 \begin{equation}
 \cov_{ij}= \frac{1}{\sum_{g_{i}}^{} \tilde{w}_{g_{i}}} \delta_{ij}~.
 \label{eq:d4}
 \end{equation}
Note that  Case--II  does not take into account the correlation between the different $\kk$ grid points that makes the off-diagonal terms of the covariance matrix to be zero. The error covariance $\cov_{ii}$ for Cases I and II match for small observation times, and they have very similar forms for very long observation times (CV) where for Case--II we have  
\begin{equation}
    \cov_{ii}=\frac{\bar{P}^2(k_i)}{N_{g_{i}}} + \frac{\Nk}{N_{g_{i}}}~\frac{\bar{T}(k_i,k_i)}{V}~.
    \label{eq:e5}
\end{equation}
This differs from the predictions for Case--I (equation \ref{eq:e4}) by the  factor $f=\Nk/N_{g_{i}}$, which appears in equation (\ref{eq:e5}). In our analysis we find that $f$ has values in the range $0.1\leq f\leq0.3$ for $k<3 \impc$ and $f\leq 1.0$ over the rest of the $k$ range considered here.  We see that the error predictions for Case--II are smaller  than those for  Case--I. The error predictions for Case--II are expected to lie somewhere in between the Gaussian predictions  and Case--I which assumes that all the $\kk$ modes in a bin are equally correlated.

We have used the resulting covariance matrices (equations \ref{eq:d3}, \ref{eq:dg3} and \ref{eq:d4}) to predict the errors for PS measurements in the different redshift slices introduced earlier in this section.

\section{Results}
{\label{sec:res}}
\begin{figure*}
    \psfrag{k}[c][c][1.5]{${k ~(\mpc^{-1})}$}
  	\psfrag{pk}[c][c][1.5]{$\quad \qquad \qquad \qquad \qquad{\Delta^2(k) ~({ \rm mK})^2}$}
  	\psfrag{PS}[c][c][1.2]{\textcolor{BlueViolet}{$\Delta_b^2(k)$}}
  	\psfrag{128}[c][c][1.2]{\textcolor{orange}{$128$}}
    \psfrag{1024}[c][c][1.2]{\textcolor{skyblue}{$1024$}}
    \psfrag{10000}[c][c][1.2]{\textcolor{forestgreen}{$10000$}}
    \psfrag{50000}[c][c][1.2]{\textcolor{gray50}{$50000$}}
    \psfrag{cv}[c][c][1.2]{\textcolor{MidnightBlue}{CV}}
    \psfrag{Non-Gaussian}[c][c][1.2]{Non-Gaussian}
    \psfrag{Gaussian}[c][c][1.2]{Gaussian}
    \psfrag{ 0.1}[c][c][1.5]{$0.1$}
    \psfrag{ 1}[c][c][1.5]{$1$}
    \psfrag{1e+03}[c][c][1.5]{$10^3$}
    \psfrag{1e+02}[c][c][1.5]{$10^2$}
    \psfrag{1e+01}[c][c][1.5]{$10^1$}
    \psfrag{1e+00}[c][c][1.5]{$10^0$}
    \psfrag{1e-01}[c][c][1.5]{$10^{-1}$}
    \psfrag{13}[c][c][1.5]{$z=13$}
    \psfrag{11}[c][c][1.5]{$z=11$}
    \psfrag{10}[c][c][1.5]{$z=10$}
    \psfrag{9}[c][c][1.5]{$z=9$}
    \psfrag{7}[c][c][1.5]{$z=7$}
    \psfrag{8}[c][c][1.5]{$z=8$}
    \centering
    \includegraphics[width=0.68\textwidth,angle=-90]{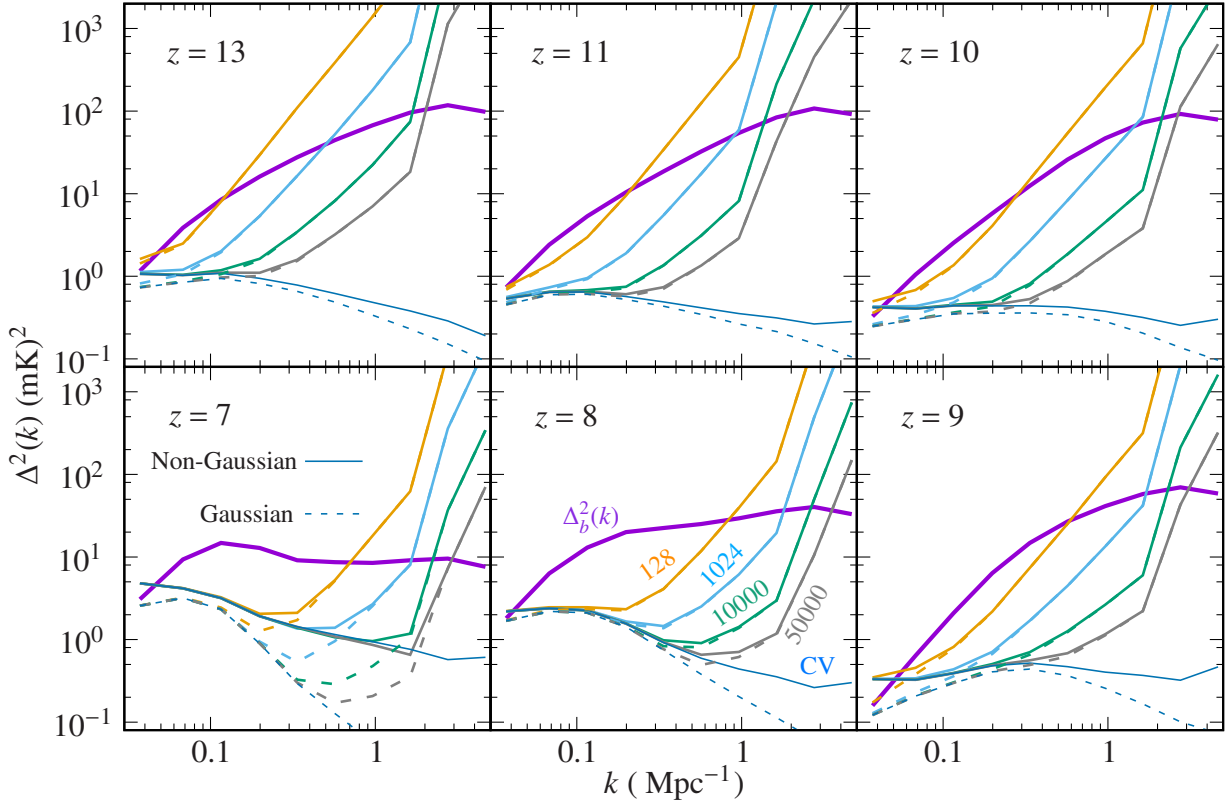}
    \caption{This shows the mean squared 21-cm brightness temperature fluctuations $\Delta_b^2(k)$ and the corresponding $5\sigma$ error estimates for different $t_{\rm obs}$ for six different redshifts considering Case--I. The solid lines represent the non-Gaussian errors and the dashed lines represent the corresponding Gaussian errors. We also show the CV that is the lowest limit of the error estimates (thin lines).}
    \label{fig:kkd_1}
\end{figure*}

\begin{figure}
  \centering
   \psfrag{km}[b][c][1.5]{$k_m~(\impc)$}
   \psfrag{z}[t][c][1.5]{$z$}
    \psfrag{128}[c][c][1.2][-25]{$t_{\rm obs}=128$}
    \psfrag{1024}[c][c][1.2]{$1024$}
    \psfrag{10000}[c][c][1.2]{$10000$}
    \psfrag{50000}[c][c][1.2]{$50000$}
   \psfrag{1e+00}[c][c][1.5]{$10^0$}
    \psfrag{1e-01}[c][c][1.5]{$10^{-1}$}
    \psfrag{ 7}[c][c][1.5]{$7$}
    \psfrag{ 8}[c][c][1.5]{$8$}
    \psfrag{ 9}[c][c][1.5]{$9$}
    \psfrag{ 10}[c][c][1.5]{$10$}
    \psfrag{ 11}[c][c][1.5]{$11$}
    \psfrag{ 12}[c][c][1.5]{$12$}
    \psfrag{ 13}[c][c][1.5]{$13$}
    \includegraphics[width=0.41\textwidth,angle=-90]{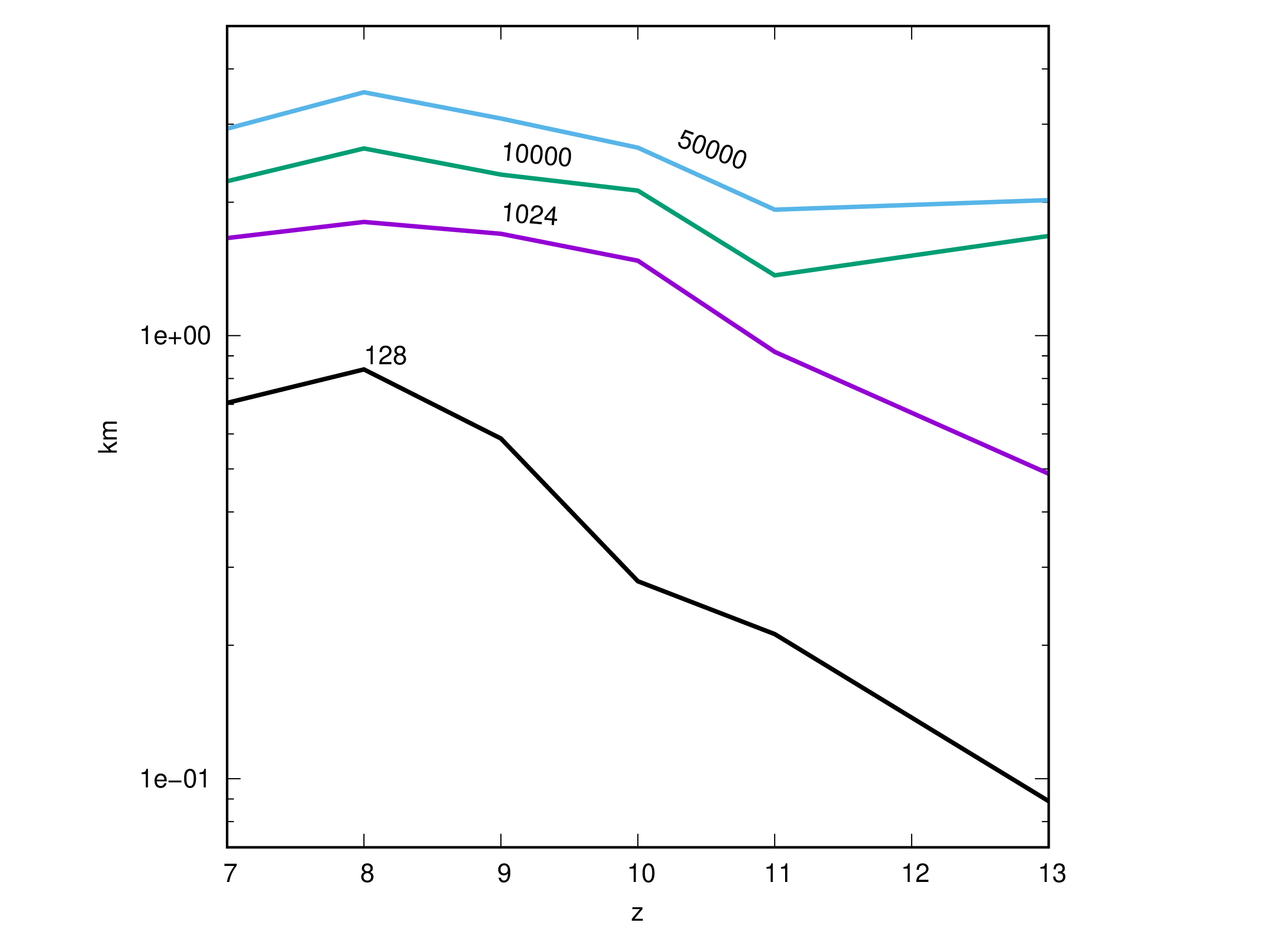}
    \caption{This shows the variation of the maximum Fourier mode $k_m$, which will be detected at a $5\sigma$ level  as a function of $z$ for the four $t_{\rm obs}$  indicated in the figure. }
    \label{fig:km}
\end{figure}

\begin{figure*}
  	\centering
 	\psfrag{k}[c][c][1.5]{${k ~(\mpc^{-1})}$}
  	\psfrag{frac}[c][c][1.5]{$\quad   \qquad \qquad \qquad \qquad \Delta \times 100 \%$}
  	\psfrag{128}[c][c][1.2]{\textcolor{orange}{$128~~$}}
  	\psfrag{1024}[c][c][1.2]{\textcolor{skyblue}{$1024~~$}}
  	\psfrag{10000}[c][c][1.2]{\textcolor{forestgreen}{$10000~~$}}
  	\psfrag{50000}[c][c][1.2]{\textcolor{gray50}{$50000~~$}}
  	\psfrag{cv}[c][c][1.2]{\textcolor{MidnightBlue}{CV~~}}
  	\psfrag{ 0.1}[c][c][1.5]{$0.1$}
  	\psfrag{ 1}[c][c][1.5]{$1$}
  	\psfrag{1e+03}[c][c][1.5]{$10^3$}
  	\psfrag{1e+02}[c][c][1.5]{$10^2$}
  	\psfrag{1e+01}[c][c][1.5]{$10^1$}
  	\psfrag{1e+00}[c][c][1.5]{$10^0$}
  	\psfrag{1e-01}[c][c][1.5]{$10^{-1}$}
  	\psfrag{13}[c][c][1.5]{$z=13$}
  	\psfrag{11}[c][c][1.5]{$z=11$}
  	\psfrag{10}[c][c][1.5]{$z=10$}
  	\psfrag{9}[c][c][1.5]{$z=9$}
  	\psfrag{7}[c][c][1.5]{$z=7$}
  	\psfrag{8}[c][c][1.5]{$z=8$}
  	\mbox{\includegraphics[width=0.59\textwidth,angle=-90]{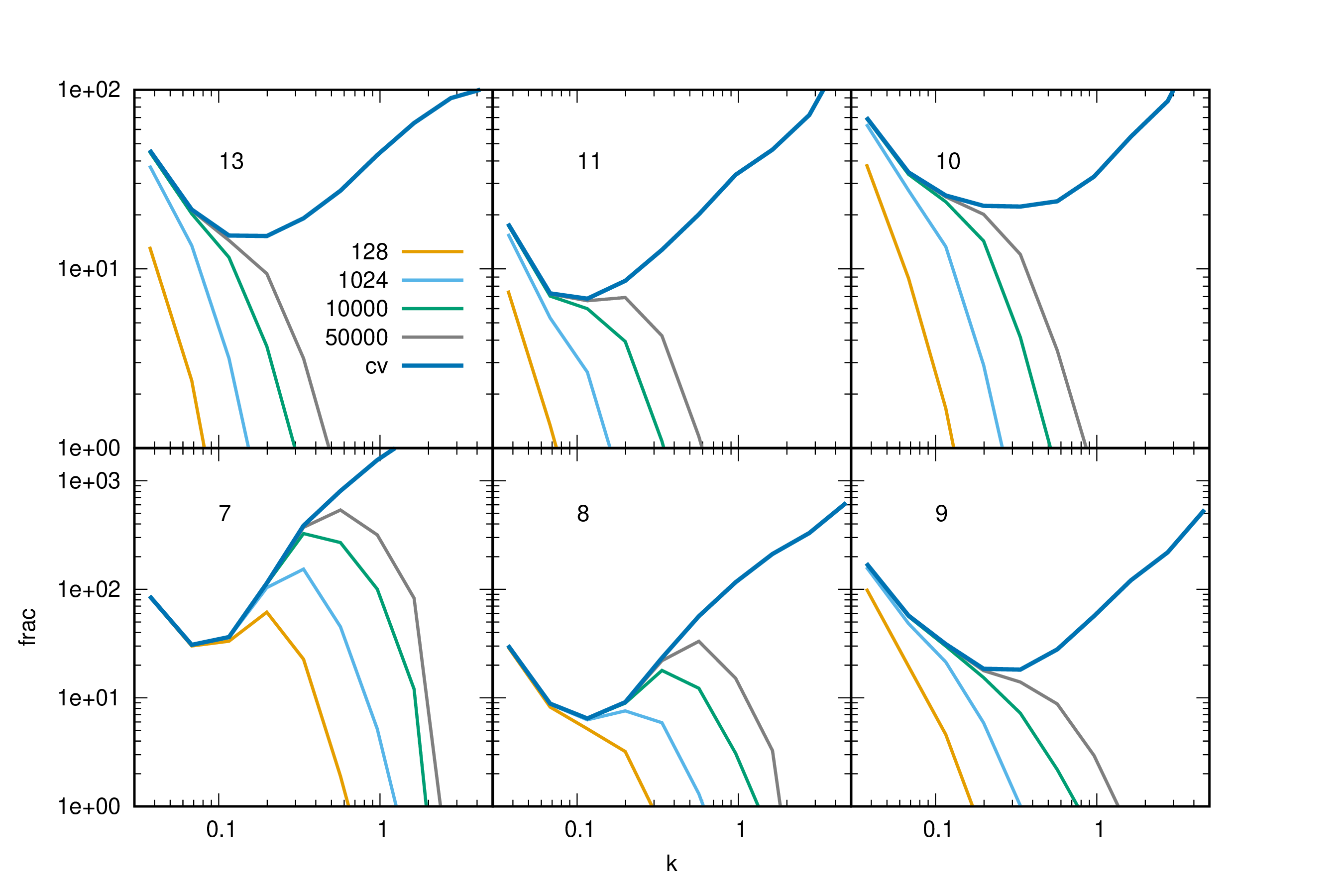}}
  	\caption{This shows the $\%$ deviation of $E_{\rm b}(k)$ with respect to the Gaussian predictions $E_{\rm bG}(k)$ considering Case--I.}
  	\label{fig:frac_1}
  \end{figure*}

  Figure \ref{fig:kkd_1} shows the dimensionless EoR 21-cm PS  $\Delta_{\rm b}^2(k)=k^3\bar{P}(k)/2\pi^2$  (solid \textcolor{BlueViolet}{purple} line)  and the corresponding $5\sigma$ error estimates  for Case--I. The solid lines represent the non-Gaussian error predictions $E_{\rm b}(k)=5\times\sqrt{\cov_{ii}}$ (equations \ref{eq:d3}) and the dashed lines represent the corresponding Gaussian error predictions $E_{\rm bG}(k)$ (equation \ref{eq:dg3}), both of these have been multiplied with $k^3/2 \pi^2$ to make them dimensionless. The error estimates have contributions from both the cosmic variance (CV) and the system noise. There are broadly two main features visible in Figure~\ref{fig:kkd_1}. (1) We see that the system noise contribution dominates the errors at large $k$. These errors come down as $t_{\rm obs}$ is increased. The errors also come down at lower $z$ where the system noise contribution is smaller  ($T_{\rm sky}$ increases with redshift).  For each $t_{\rm obs}$ and $z$ we can identify a largest mode ($k_m$) below which $(k \le k_m)$  a $5 \sigma$ detection of the 21-cm power spectrum will be possible. A larger $k$ range becomes accessible for a $5 \sigma$ detection ($k_m$ increases) as  $t_{\rm obs}$ is increased or we move to a lower $z$. This is studied in more detail in Figure \ref{fig:km}, which we discuss later. (2) We see noticeable differences between $E_{\rm b}(k)$ and $E_{\rm bG}(k)$. These differences are most prominent for the CV predictions that correspond to the limit $t_{\rm obs}\rightarrow \infty$, where the system noise becomes insignificant. The system noise contribution is inherently Gaussian, whereas the 21-cm signal is non-Gaussian. We find that the values of  $E_{\rm b}(k)$ and $E_{\rm bG}(k)$ match for small $t_{\rm obs}$ when the system noise dominates the errors. The differences between $E_{\rm b}(k)$ and $E_{\rm bG}(k)$  become noticeable as $t_{\rm obs}$ is increased. The differences are primarily noticeable at small $k$ where there is a relatively smaller system noise contribution as compared to large $k$. The differences also become more pronounced as we move to lower $z$, where there is a smaller system noise contribution. The differences between  $E_{\rm b}(k)$ and $E_{\rm bG}(k)$ are studied in detail in Figure \ref{fig:frac_1}, which we discuss later.

Considering  Figure \ref{fig:kkd_1}, we see that the predicted error estimates  $E_{\rm b}(k)$ all increase with $k$ mainly due to the system noise contribution in contrast to the expected signal $\Delta_{\rm b}^2(k)$, which is relatively flat across the relevant $k$ range. This implies that for any given $t_{\rm obs}$ a detection of the signal will only be possible at small $k$ whereas the errors in the power spectrum will dominate at large $k$. Figure \ref{fig:km} shows the largest $k$ mode $k_m$, below which SKA-Low will be able to measure the EoR 21-cm PS at $\geq 5 \sigma$ confidence. We show this as a function of $z$ for the four representative values of $t_{\rm obs}$ indicated in the figure. We see that the value of $k_m$ increases as $z$ decreases {\it i.e.} for a fixed observation time, we will progressively be able to probe a larger range of length-scales as reionization progresses. This is primarily a consequence of the fact that the system noise comes down at lower $z$, further the amplitude of the 21-cm PS also increases as reionization progresses. However, the amplitude peaks at $\sim 50\%$ reionization and drops beyond this, causing $k_m$ to fall at $z=7$.  Considering $t_{\rm obs}=128$ hours we find that there is a limited $k$ range across which a $5\sigma$ detection of the 21-cm PS is possible. This is restricted to $k \le 0.2 \impc$ at high $z$ $(=11,13)$ and increases somewhat to $k \le 0.8 \impc$ at $z=7$ and $8$.  There is a significant increase in the values of $k_m$ (by a factor of $\sim 2.5-5$) if $t_{\rm obs}$ is increased to $1024$ hours. We see that with $t_{\rm obs}=1024$ hours a $5\sigma$ detection will be possible in the range $ k \le 1 \impc$  at $z \le 11$. The value of $k_m$ increases gradually if $t_{\rm obs}$ is increased beyond $1024$ hours. However, we see an exception at $z=13$ where there is a significant increase in $k_m$ if $t_{\rm obs}$ is increased beyond $1024$ hours.  The values of $k_m$ increases very slowly for $t_{\rm obs} \geq 10000$ hours and $k_m$ values are in the range $2-4 \, \impc$ for $t_{\rm obs} = 50000$ hours.


\begin{figure*}
  	\psfrag{Case--I}[c][c][1.2]{Case--I}
   \psfrag{Case--II}[c][c][1.2]{Case--II}
   \psfrag{Gaussian}[c][c][1.2]{Gaussian}
   \psfrag{diff}[c][c][1.5]{$\Delta$}
   \psfrag{Gaussian CV}[c][c][1.2]{Gaussian CV}
   \psfrag{Non-Gaussian CV}[c][c][1.2]{Non-Gaussian CV}
   \psfrag{SNR}[c][c][1.5]{ \qquad \qquad \qquad \qquad \qquad SNR}
   \psfrag{del}[c][c][1.5]{\qquad \qquad \qquad \qquad \qquad $\Delta \times 100 \%$}
   \psfrag{t}[c][c][1.5]{$t_{\rm obs}$ (hours)}
   \psfrag{2}[c][c][1.5]{$10^2$}
    \psfrag{3}[c][c][1.5]{$10^3$}
    \psfrag{4}[c][c][1.5]{$10^4$}
    \psfrag{ 10}[c][c][1.5]{$10^1$}
    \psfrag{10}[c][c][1.5]{$10^1$}
    \psfrag{1}[c][c][1.5]{$10^0$}
    \psfrag{1e+04}[c][c][1.5]{$10^4$}
    \psfrag{1e+05}[c][c][1.5]{$10^5$}
    \psfrag{1e+03}[c][c][1.5]{$10^3$}
    \psfrag{1e+02}[c][c][1.5]{$10^2$}
    \psfrag{1e+01}[c][c][1.5]{$10^1$}
    \psfrag{k=0.037 Mpc-1}[c][c][1.2]{$k_1=0.04 \impc$}
    \includegraphics[width=0.59\textwidth,angle=-90]{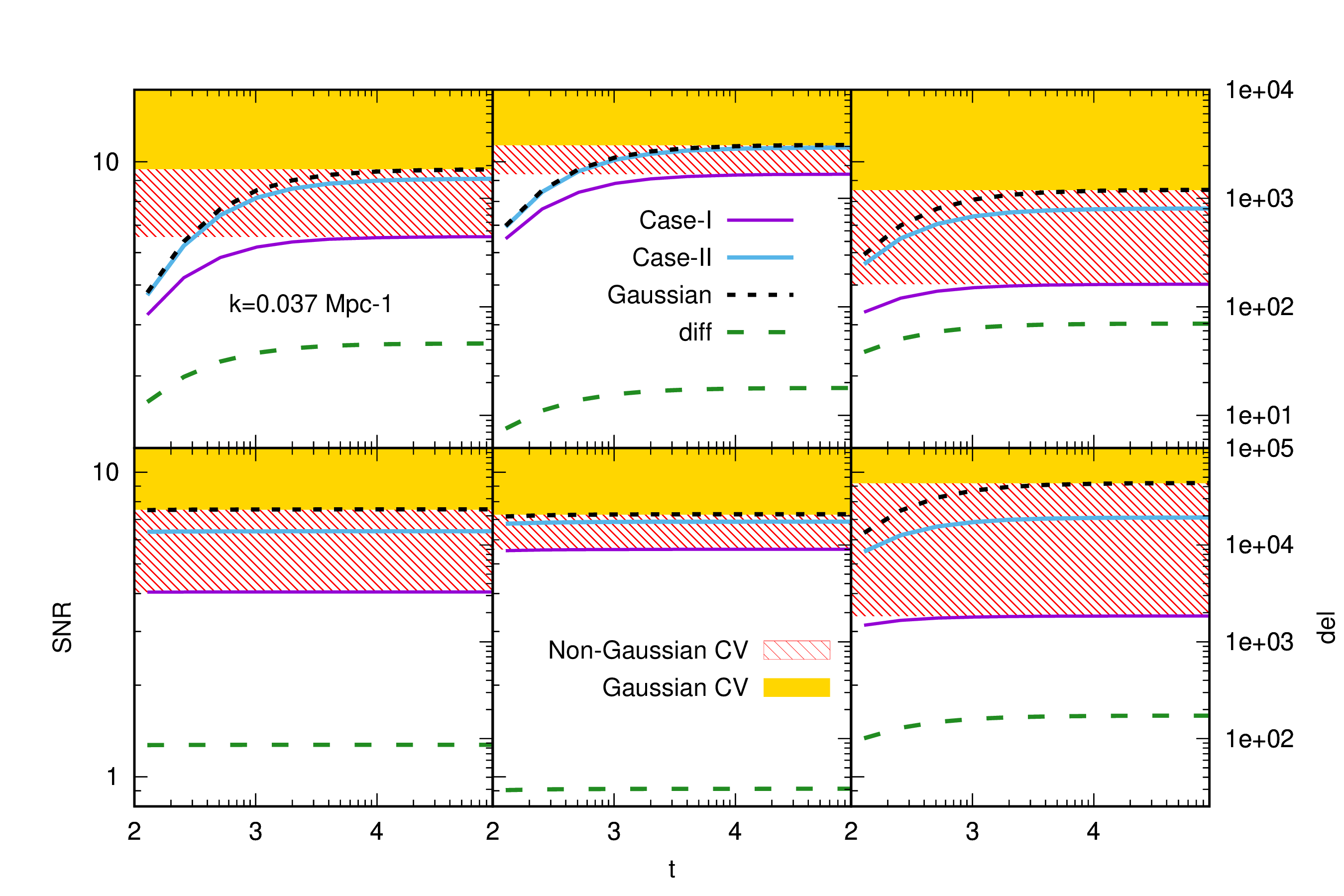}
    \caption{This shows the SNR (left axis) as a function of the observing time $t_{\rm obs}$    for $k=0.04 \impc$, which is representative of large length-scales.  Results are shown for Case--I, Case--II and the Gaussian predictions, while the two shaded regions  demarcate the CV limits for Case--I and the Gaussian predictions, respectively. The dashed line (green) shows $\Delta$ (right axis) as a function of $t_{\rm obs}$.  The different panels,  each of which corresponds to a different redshift, are arranged the same way  as in Figure \ref{fig:frac_1}.}
    \label{fig:SNR_1}
\end{figure*}
 
   Figure \ref{fig:frac_1} shows the deviation $\Delta=(E_b-E_{bG})/E_{bG}$ of the non-Gaussian error estimates with respect to the corresponding Gaussian estimates. These deviations arise due to  the contribution from the trispectrum (equation ~\ref{eq:d3}). Earlier studies \citep{Mondal_I,Mondal_II} show that  the trispectrum increases at larger $k$  (smaller length-scales), and it also increases as reionization proceeds {\it i.e.} $z$ decreases. These effects are reflected in the behaviour of the CV, which ignores the system noise. Considering the CV, we see that the deviations are minimum at around $k_{\rm min} \sim 0.1 - 0.3 \, \impc$, and  the deviations increase monotonically at both smaller and larger $k$ values. At the smallest $k$ bin ($0.04 \impc$) we find $\Delta \ge 100 \%$ at $z=7$ and $9$, whereas $\Delta \sim 20 \%$ to $50 \%$ for the other redshifts. The values of $\Delta$ increase  significantly at $k>k_{\rm min}$ with deviations of order $\sim 100 \%$ or larger at $k \approx 4 \, \impc$ for the entire $z$ range.  Considering the redshift evolution of CV, we see that at large $k$ the deviations from the Gaussian predictions increase as reionization proceeds.

   We see that for $k < k_{min}$ the values of $\Delta$ approach the CV limit within $t_{\rm obs}=1024$ hours for $z\ge 9$ and within $t_{\rm obs}=128$ hours for lower redshifts. We find that the bins at $k>k_{\rm min}$ are largely system noise dominated, and the deviations at these bins are small for $z\leq 9$ even for an observing time of $50000$ hours. However, at $z=8$ we find that $\Delta$ also increases at large $k$ ($>k_{\rm min}$) for $t_{\rm obs}\ge 10000$ hours and we have $\Delta \sim 40 \%$ at $k \sim 0.5 \impc$ for  $t_{\rm obs} = 50000$ hours. These deviations increase significantly at $z=7$, where $\Delta \ge 100 \%$ at $k \sim 0.2-0.5 \impc$ for $t_{\rm obs}=1024$ hours. The $k$ range where $\Delta \ge 100 \%$ increases further to $k \sim 0.2 - 1 \impc$ if $t_{\rm obs}$ is increased further to $10000$ hours.

\begin{figure*}
   \psfrag{Case--I}[c][c][1.2]{Case--I}
   \psfrag{Case--II}[c][c][1.2]{Case--II}
   \psfrag{Gaussian}[c][c][1.2]{Gaussian}
   \psfrag{diff}[c][c][1.2]{$\Delta$}
   \psfrag{Gaussian CV}[c][c][1.2]{Gaussian CV}
   \psfrag{Non-Gaussian CV}[c][c][1.2]{Non-Gaussian CV}
   \psfrag{SNR}[c][c][1.5]{ \qquad \qquad \qquad \qquad \qquad SNR}
   \psfrag{del}[c][c][1.5]{\qquad \qquad \qquad \qquad \qquad $\Delta \times 100 \%$}
   \psfrag{t}[c][c][1.5]{$t_{\rm obs}$ (hours)}
   \psfrag{2}[c][c][1.5]{$10^2$}
    \psfrag{3}[c][c][1.5]{$10^3$}
    \psfrag{4}[c][c][1.5]{$10^4$}
    \psfrag{5}[c][c][1.5]{$10^5$}
    \psfrag{1e+04}[c][c][1.5]{$10^4$}
    \psfrag{1e+03}[c][c][1.5]{$10^3$}
    \psfrag{1e+02}[c][c][1.5]{$10^2$}
    \psfrag{1e+01}[c][c][1.5]{$10^1$}
    \psfrag{ 100}[c][c][1.5]{$10^2$}
    \psfrag{ 10}[c][c][1.5]{$10^1$}
    \psfrag{ 1}[c][c][1.5]{$10^0$}
    \psfrag{1}[c][c][1.5]{$10^0$}
    \psfrag{10}[c][c][1.5]{$10^1$}
    \psfrag{100}[c][c][1.5]{$10^2$}
    \psfrag{k=0.569 Mpc-1}[c][c][1.2]{$k_6=0.57 \impc$}
    \includegraphics[width=0.62\textwidth,angle=-90]{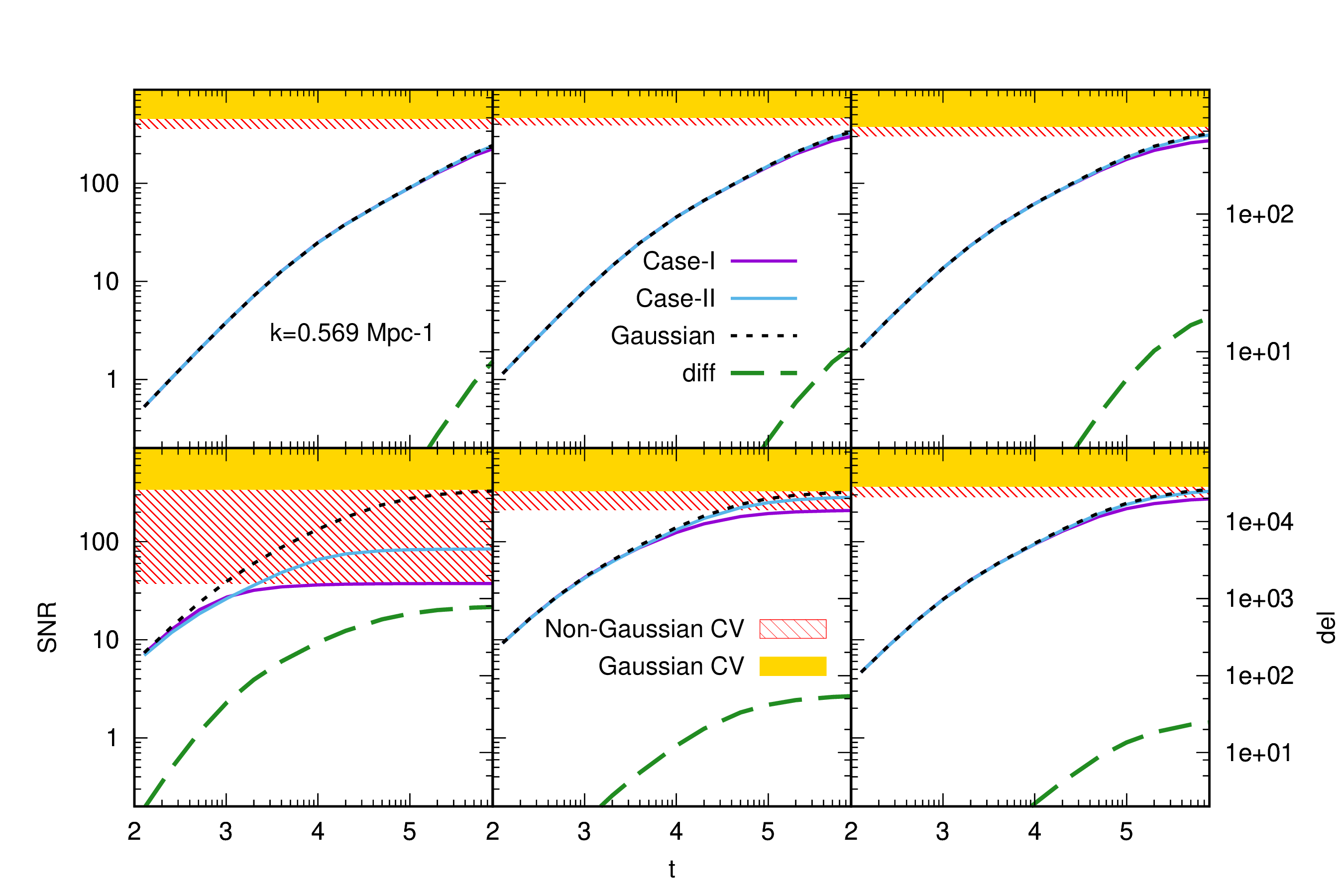}
    \caption{Same   as Figure \ref{fig:SNR_1} for  $k=0.57 \impc$.}
    \label{fig:SNR_6}
\end{figure*}

\begin{figure*}
  	\psfrag{Case--I}[c][c][1.2]{Case--I}
   \psfrag{Case--II}[c][c][1.2]{Case--II}
   \psfrag{Gaussian}[c][c][1.2]{Gaussian}
   \psfrag{diff}[c][c][1.5]{$\Delta$}
   \psfrag{Gaussian CV}[c][c][1.2]{Gaussian CV}
   \psfrag{Non-Gaussian CV}[c][c][1.2]{Non-Gaussian CV}
   \psfrag{SNR}[c][c][1.5]{ \qquad \qquad \qquad \qquad \qquad SNR}
   \psfrag{del}[c][c][1.5]{\qquad \qquad \qquad \qquad \qquad $\Delta \times 100 \%$ }
   \psfrag{t}[c][c][1.5]{$t_{\rm obs}$ (hours)}
   \psfrag{2}[c][c][1.5]{$10^2$}
    \psfrag{3}[c][c][1.5]{$10^3$}
    \psfrag{4}[c][c][1.5]{$10^4$}
    \psfrag{5}[c][c][1.5]{$10^5$}
    \psfrag{ 0.01}[c][c][1.5]{$10^{-2}$}
    \psfrag{ 0.1}[c][c][1.5]{$10^{-1}$}
    \psfrag{ 1}[c][c][1.5]{$10^0$}
    \psfrag{ 10}[c][c][1.5]{$10^1$}
    \psfrag{ 100}[c][c][1.5]{$10^2$}
    \psfrag{ 1000}[c][c][1.5]{$10^3$}
    \psfrag{1}[c][c][1.5]{$10^0$}
    \psfrag{100}[c][c][1.5]{$10^2$}
    \psfrag{1000}[c][c][1.5]{$10^3$}
    \psfrag{1e+04}[c][c][1.5]{$10^4$}
    \psfrag{1e+03}[c][c][1.5]{$10^3$}
    \psfrag{1e+02}[c][c][1.5]{$10^2$}
    \psfrag{1e+01}[c][c][1.5]{$10^1$}
    \psfrag{1e+00}[c][c][1.5]{$10^0$}
    \psfrag{1e-01}[c][c][1.5]{$10^{-1}$}
    \psfrag{k=1.626 Mpc-1}[c][c][1.2]{$k_8=1.63 \impc$~~}
    \includegraphics[width=0.62\textwidth,angle=-90]{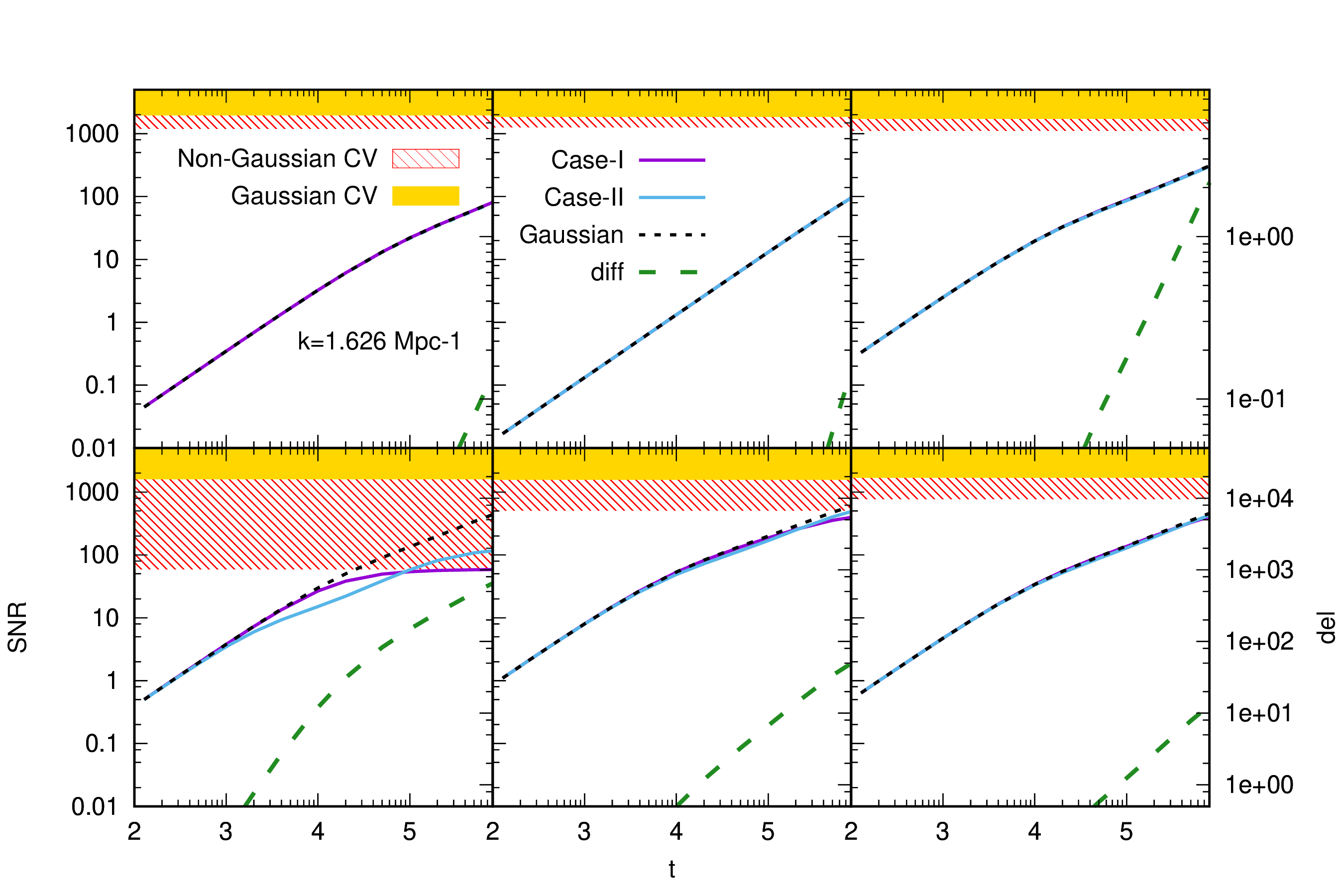}
    \caption{Same as Figure \ref{fig:SNR_1}  for $k=1.63 \impc$.}
    \label{fig:SNR_8}
\end{figure*}

 We next consider how the SNR for the 21-cm PS grows with increasing observation time $t_{\rm obs}$. Figures \ref{fig:SNR_1}--\ref{fig:SNR_8} show  the results for three  representative $k$ bins located at $0.04 \impc$ (large scales), $0.57 \impc$ (intermediate scales) and $1.63 \impc$ (small scales), respectively. The SNR values are shown for both Case--I (purple solid line) and Case--II (blue solid line), as well as the Gaussian predictions (dotted black line). The CV limits ($t_{\rm obs} \rightarrow \infty$) are shown as shaded regions for both the non-Gaussian (Case--I) and Gaussian predictions. We find that the differences between Case--I, II and the Gaussian predictions are noticeable only when the SNR approaches the CV limit. The Gaussian predictions are the most optimistic of the three, and the SNR values for Case--II are typically between those for Case--I and the Gaussian predictions. The figure also shows how $\Delta$ increases with $t_{\rm obs}$ at the specified values of $k$.


 Considering  the lowest $k$ bin ($k=0.04 \impc$; Figure \ref{fig:SNR_1}), the SNR is largely constrained by the CV with a relatively small system noise contribution. The   SNR saturates to the  CV limit within a few hundred hours of observations at $z\leq 10$ and within $t_{\rm obs}\sim 3000$ hours for $z>10$. Considering Case--I, a $\geq 5\sigma$ measurement  of the EoR 21-cm PS will be possible with $t_{\rm obs}\geq 128$ hours at redshifts $z=13,11,8$ and with $t_{\rm obs}\geq 3000$ hours at $z=10$,  whereas a $5\sigma$ detection is limited by the CV at $z=7$ and $9$. However, the Case--II predictions are more optimistic and they predict a $5 \sigma$ detection to be possible.  The deviations between the non-Gaussian and Gaussian predictions are found to become important $(\Delta \ge 50\%)$ within a few hundred hours of observations at redshifts $z=10,9~{\rm and}~7$.

Considering $k=0.57 \impc$  (Figure \ref{fig:SNR_6}), the limiting SNR (CV) increases to  values $> 100$ at $z \ge 8$ and $\sim 40$ at $z=7$, implying that a high-precision measurement of the EoR 21-cm PS is possible at these length-scales provided that $t_{\rm obs}$ is adequately large. The $t_{\rm obs}$ needed for a $5 \sigma$ detection is $\sim 1000$ hours at $z=13$ and it comes down at lower $z$ to $\sim 128$ hours at  $z=8$ and $7$. The SNR is highest at $z=8$ and we have SNR $\approx 100$ in $\sim 4000$ hours of observations. The non-Gaussian effects make a relatively small contribution to the error predictions at this length-scale with $\Delta \leq 20 \%$ in the range $z \geq 8$ for $t_{\rm obs} \le 10^4$ hours. The non-Gaussian effects increase somewhat at $z=7$, where we have $\Delta \approx 250 \%$ for $t_{\rm obs} \approx 10^4$ hours.

 Considering the  bin at $k=1.63 \impc$ (Figure \ref{fig:SNR_8}) the SNR is largely system noise dominated. The SNR is well below the cosmic variance limit and increases with $t_{\rm obs}$ for the range shown in the figure except for the Case--I at $z=7$.  A $5\sigma$ detection will be possible with $t_{\rm obs}\approx 20000,~40000,~2000,~1000~{\rm and}~600$ hours at $z=13,11,10,9,~{\rm and}~8$, respectively.  The value of the 21-cm PS falls at $z=7$ and the   minimum observation time required for a $5\sigma$ detection increases to $1,000$ hours. The inherent non-Gaussianity of the 21-cm signal is important  only at $z=7$, where we have $10\% \leq \Delta \leq 100\%$ for $10^4 ~{\rm hours}~ \leq t_{\rm obs} < 10^5$ hours.


   \begin{figure*}
   	\centering 
   	\psfrag{k}[c][b][1.5]{${k ~(\impc)}$}
   	\psfrag{k1}[c][c][1.5]{$\qquad \qquad \qquad \qquad{k ~(\impc)}$}
   	\psfrag{r}[c][c][1.5]{$r_{ij}$}
   	\psfrag{1.0}[c][c][1.2]{$1.0$}
   	\psfrag{0.8}[c][c][1.2]{$0.8$}
   	\psfrag{0.6}[c][c][1.2]{$0.6$}
   	\psfrag{0.4}[c][c][1.2]{$0.4$}
   	\psfrag{0.2}[c][c][1.2]{$0.2$}
   	\psfrag{0.0}[c][c][1.2]{$0.0$}
   	\psfrag{-0.2}[c][c][1.2]{$-0.2$}
   	\psfrag{-0.4}[c][c][1.2]{$-0.4$}
   	\psfrag{-0.6}[c][c][1.2]{$-0.6$}
   	\psfrag{0.04}[c][c][1.2]{$0.04$}
   	\psfrag{0.07}[c][c][1.2]{$0.07$}
   	\psfrag{0.12}[c][c][1.2]{$0.12$}
   	\psfrag{0.20}[c][c][1.2]{$0.20$}
   	\psfrag{0.34}[c][c][1.2]{$0.34$}
   	\psfrag{0.57}[c][c][1.2]{$0.57$}
   	\psfrag{0.96}[c][c][1.2]{$0.96$}
   	\psfrag{1.63}[c][c][1.2]{$1.63$}
   	\psfrag{2.75}[c][c][1.2]{$2.75$}
   	\psfrag{4.66}[c][c][1.2]{$4.66$}
   	
   	\mbox{\includegraphics[width=0.64\textwidth,angle=-90]{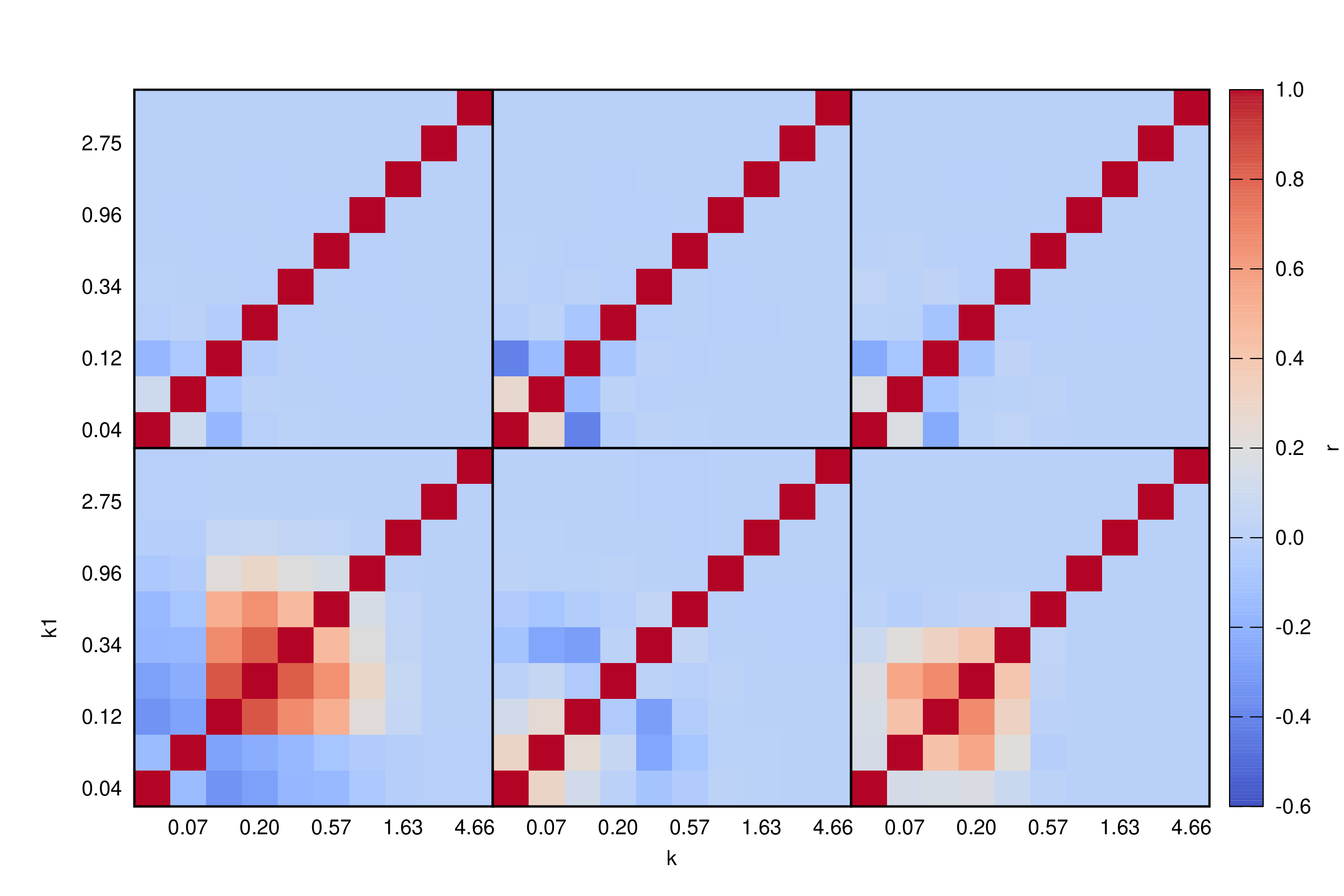}}
   	\caption{This shows the correlation coefficient $r_{ij}$ for  the errors at different $k$ bins for $1024$ hours of observations.
   	The different panels,  each of which corresponds to a different redshift, are arranged the same way  as in Figure \ref{fig:frac_1}.
   	}
    \label{fig:cov_1024}
   \end{figure*}

We now discuss the off-diagonal elements of the covariance matrix $\cov_{ij}$, which is a measure of the correlation between error estimates at different $k$ bins. The off-diagonal terms of the covariance $\cov_{ij}$ do not change with the observation time as we see in equation (\ref{eq:d3}). It is convenient to consider  the  dimensionless correlation coefficients $r_{ij}=\cov_{ij}/\sqrt{\cov_{ii}\cov_{jj}}$. The value  $r_{ij}=1$ indicates a perfect  correlation between the errors at the two bins, whereas $r_{ij}=-1$ implies a complete  anticorrelation. The errors in the two bins are completely uncorrelated if $r_{ij}=0$ {\it i.e.} the two PS measurements are independent. Values $r_{ij}>0$ and $r_{ij}<0$  indicate partial correlation and anticorrelation, respectively. An earlier work \citep{Mondal_II} presents a detailed analysis of the correlations $r_{ij}$ evaluated from simulations. It was found that the non-Gaussianity inherent in the EoR 21-cm signal introduces a complex pattern of correlations  and anticorrelations between the different $k$ bins. It was further found that these correlations (and anticorrelations) were statistically significant, {\it i.e.} they were in excess of the statistical fluctuations expected if the signal were purely a Gaussian random field. However, the earlier work did not include the effects of the baseline sampling and system noise corresponding to observations with a radio-interferometric array. For an array like SKA-Low, the correlation coefficient $r_{ij}$ is dependent on the observation time through the diagonal elements $\cov_{ii}$, which appear in the denominator. As discussed earlier, the values of $\cov_{ii}$  are typically large for small $t_{\rm obs}$ where they are system noise dominated. The relative significance of the correlations between the errors in different $k$ bins is small for small $t_{\rm obs}$ where $r_{ij}$ has small values. The relative significance of these correlations increases as $\cov_{ii}$ approaches the CV and we have considered $t_{\rm obs}=1024$ hours for our analysis. The values of $r_{ij}$ will  increase if we consider a larger observation time.

Considering Figure \ref{fig:cov_1024}, we see that in addition to $r_{ii}=1$ (by definition) for all the diagonal elements,  we have  both positive and negative values of $r_{ij}$. The redshifts $z=13,~11$ and $10$ show very similar features with a positive correlation ($r_{ij} \sim 0.1-0.3$) between the two smallest $k$ bins ($0.04,~0.07 \, \impc$), and the third bin  ($0.12 \, \impc$) is anticorrelated ($r_{ij}\sim -0.4$ to $-0.1$) with the two smaller $k$ bins and one larger $k$ bin ($0.20 \, \impc$). The nature of these correlations changes at $z=9$, where the first five $k$ bins ($k\leq 0.34 \impc$) are correlated. Of these, the four largest $k$ bins are strongly correlated  ($0.2 \le r_{ij} < 0.7$) among themselves whereas the smallest $k$ bin is only mildly correlated ($r_{ij}<0.2$) with the other bins. At $z=8$, the first three $k$ bins are  correlated ($r_{ij}\leq 0.3$) whereas the fifth bin shows anticorrelations ($r_{ij}>-0.3$) with the second and third bins. Considering $z=7$, the first two $k$ bins are anticorrelated ($r_{ij}\geq -0.3$) with the other bins while the next five $k$ bins show strong correlations ($0.15\leq r_{ij}\leq 0.85$). We thus see that there are noticeable  correlations and anticorrelations between the errors in the estimated 21-cm PS  in different $k$ bins at all stages of reionization. These correlations span a wide range of $k$ modes depending on the redshift.

\section{Effects of Foregrounds}\label{sec:fgd}
\begin{figure*}
    \centering
    \psfrag{k}[c][c][1.5]{$\qquad \qquad \qquad \qquad \qquad{k ~(\impc)}$}
    \psfrag{SNR}[c][c][1.5]{SNR \qquad \qquad \qquad \qquad}
    \psfrag{z}[c][b][1.5]{$z$}
    \psfrag{ 0.1}[c][c][1.2]{$10^{-1}$}
    \psfrag{ 1}[c][c][1.2]{$10^{0}$}
    \psfrag{ 7}[c][c][1.2]{$7$}
    \psfrag{ 8}[c][c][1.2]{$8$}
    \psfrag{ 9}[c][c][1.2]{$9$}
    \psfrag{ 10}[c][c][1.2]{$10$}
    \psfrag{ 11}[c][c][1.2]{$11$}
    \psfrag{ 12}[c][c][1.2]{$12$}
    \psfrag{ 13}[c][c][1.2]{$13$}
    \psfrag{3}[c][c][0.8]{$\mathbf{3}$}
    \psfrag{5}[c][c][0.8]{$\mathbf{5}$}
    \psfrag{10}[c][c][0.8]{$\mathbf{10}$}
    \psfrag{30}[c][c][0.8]{$\mathbf{30}$}
    \psfrag{50}[c][c][0.8]{$\mathbf{50}$}
    \psfrag{Optimistic}[c][c][1.0]{\textcolor{yellow}{\bf{Optimistic}}}
    \psfrag{Moderate}[c][c][1.0]{\textcolor{black}{\bf{Moderate}}}
    \psfrag{Pessimistic}[c][c][1.0]{\textcolor{black}{\bf{Pessimistic}}}
    \psfrag{ 0}[c][c][1.2]{$0$}
    \psfrag{ 10}[c][c][1.2]{$10$}
    \psfrag{ 20}[c][c][1.2]{$20$}
    \psfrag{ 30}[c][c][1.2]{$30$}
    \psfrag{ 40}[c][c][1.2]{$40$}
    \psfrag{ 50}[c][c][1.2]{$50$}
    \psfrag{ 60}[c][c][1.2]{$60$}
    \psfrag{ 80}[c][c][1.2]{$80$}
    \psfrag{ 100}[c][c][1.2]{$100$}
    \includegraphics[scale=0.7, angle=-90]{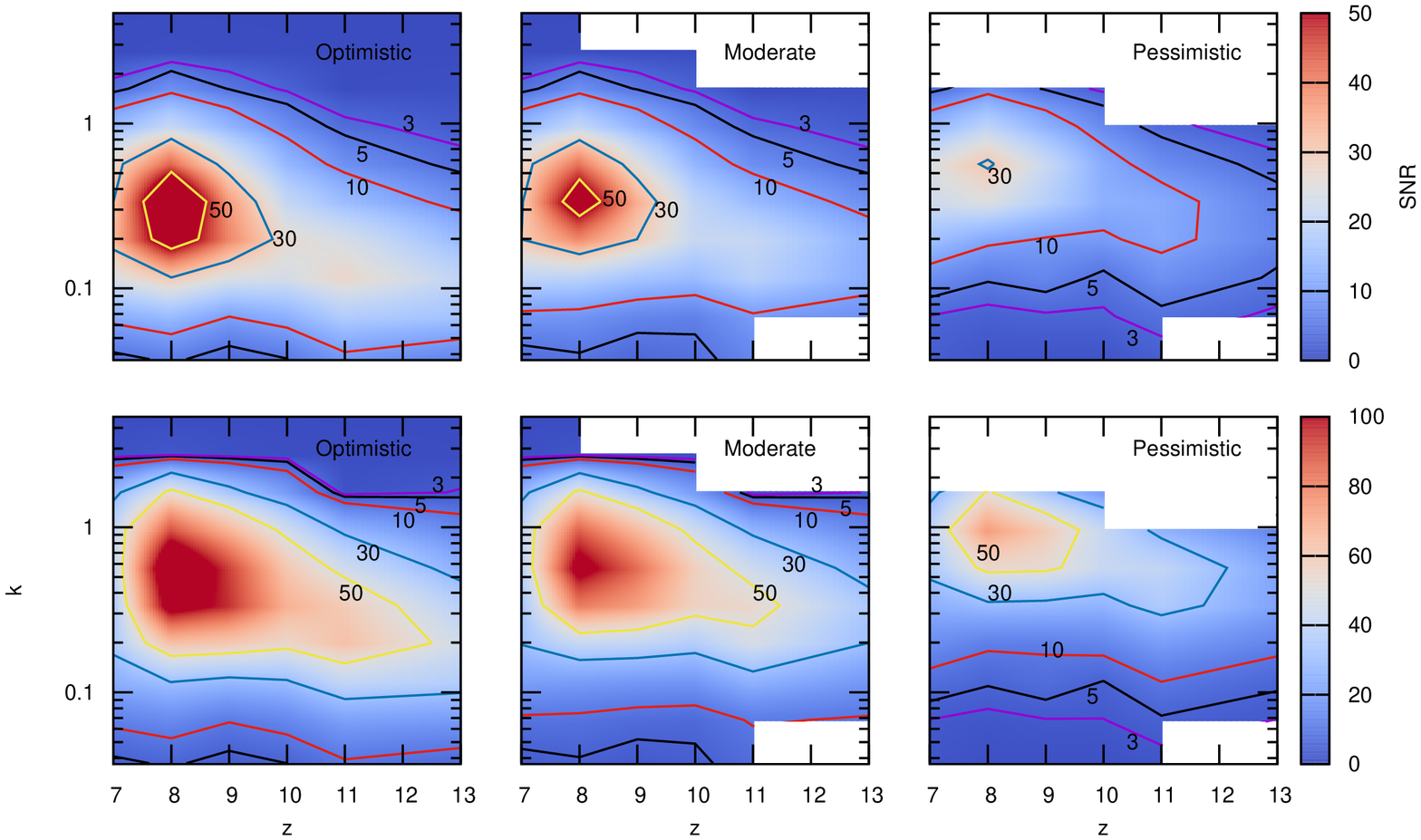}
    \caption{This shows the predicted SNR (non-Gaussian Case--I) as a function of $k$ and $z$ for the three different foreground scenarios. The top and bottom panels show the results for $1024$ hours and $10000$ hours of observations respectively. Note that the color bars are different for the top  and bottom panels.}
    \label{fig:SNR_wedge}
\end{figure*}

Foregrounds, which are almost $4-5$ order magnitude larger than the \signal (e.g. \citealt{Abhik_2012}), are a major challenge for measuring the EoR 21-cm PS. There are several approaches that have been proposed to handle the foreground problem, one of these being foreground removal  (e.g. \citealt{Morales_2006,Ali_2008,Harker_2009,Parsons_2012b,Bonaldi_2015,Chapman_2015,Pober_2016b}). The entire analysis until now has assumed that the foregrounds have been perfectly modelled and removed, following \citet{SumanCh_2019} we refer to this as  as the ``Optimistic'' scenario in the subsequent discussion.

The foreground contribution to the 21-cm PS is predicted to be localized within a wedge in the ($\kk_{\perp},k_\parallel$) plane \citep{Datta_2010}, the boundary of this wedge being defined through \citep{Morales_2012} 
\begin{equation}
    k_{\parallel} = \left[\frac{r_c ~\sin(\theta_{\rm L})}{r^\prime_c ~\nu_c} \right] k_{\perp}
    \label{eq:wedge}
\end{equation}
where $\theta_{\rm L}$ is the maximum angular position in the sky (relative to the telescope pointing) from which foregrounds contaminate the signal. The $\kk(\kk_{\perp},k_\parallel)$ modes outside this foreground wedge are expected to be free of foreground contamination, and the `foreground avoidance' technique (e.g. \citealt{Pober_2013,Kerrigan_2018}) proposes to utilize only these modes to estimate the EoR 21-cm PS. Typically $\theta_{\rm L}=90^{\circ}$ corresponding to the horizon that is the maximum angle from which the foregrounds contaminate the signal.  However, it is possible to  taper the telescope's field of view 
\citep{Abhik_2011,Choudhuri_2016} and thereby restrict $\theta_{\rm L}$ to an angle smaller than the horizon. Here, in addition to $\theta_{\rm L}=90^{\circ}$ we also consider a situation in which we assume that tapering is used whereby $\theta_{\rm L}=3\times {\rm FWHM}/2$
where ${\rm FWHM}$ is the Full Width Half Maxima of the SKA-Low primary beam. Note that ${\rm FWHM}$ changes with frequency and it  is $\sim 6^{\circ}$ at $z=8$. 
Following \citet{SumanCh_2019}, we refer to the two cases $\theta_{\rm L}=3\times {\rm FWHM}/2$ and $90^{\circ}$
as the `Moderate' and `Pessimistic' scenarios, respectively.

Figure \ref{fig:SNR_wedge} shows the SNR for detecting the EoR 21-cm PS at different $k$ bins for various $z$ values  considering the non-Gaussian error covariance for Case--I. Starting from the left, the three columns show the predictions for the Optimistic, Moderate and Pessimistic scenarios, respectively, while the upper and lower rows correspond to $t_{\rm obs}=1024$ and $10000$ hours respectively. The first point to note is that a few  $k$ bins for which all the $k$ modes are within the foreground wedge  are excluded from the  detection of the EoR 21-cm PS. These excluded $k$ bins  occur at the two extremities (large $k$ and small $k$).
Further in equation (\ref{eq:wedge}) the factor $r_c /(r^\prime_c ~\nu_c) \sim \sqrt{1+z}$ 
 causes the extent of the foreground  wedge to increase with $z$  ($\theta_{\rm L}$ also increases with $z$ in the Moderate scenario) and we see that the extent of the excluded $k$ bins increases at higher redshifts.

 In each $k$ bin the number of $k$ modes  that can be used for measuring the 21-cm PS  decreases from the the Optimistic to the Moderate and then the Pessimistic scenarios.  This causes the 
  SNR to decrease from the Optimistic to the Moderate scenario, and the SNR  decreases even further for the  Pessimistic scenario. The $k$  range where the SNR exceeds $5$ does not change very much from the Optimistic to Moderate scenario for both $1024$ and $10000$ hours, except for a small raising of the lower $k$ limit.  
  The lower $k$ limit for a $5 \sigma$ detection increases significantly for the Pessimistic scenario, however the upper $k$ limit is not much affected outside the excluded bins. In all cases the SNR peaks at $z=8$.  Considering the region where the SNR exceeds $30$, we see that for the Optimistic scenario with $1024$ hours this spans  from $z=7-10$ and $k=0.1 \, \impc$ to $0.8 \, \impc$.  The  range  shrinks to $z=7-9$ and $k=0.2-0.8 \, \impc$ for the Moderate scenario and shrinks even further to a very small region around $z=8$ and $k=0.6\, \impc$ for the Pessimistic scenario. The range where the SNR exceeds $30$ increases significantly if the observing time is increased to $10000$ hours, this is particularly prominent for the Pessimistic scenario  where both the $z$ and $k$ ranges  are considerably increased compared to $1024$ hours. 
 \begin{figure*}
    \centering
    \psfrag{k}[c][b][1.5]{$\qquad \qquad \qquad \qquad \qquad{k ~(\impc)}$}
    \psfrag{NG/G}[c][t][1.5]{$\Delta (\%)$ \qquad \qquad \qquad \qquad}
    \psfrag{z}[c][b][1.5]{$z$}
    \psfrag{ 0.1}[c][c][1.2]{$10^{-1}$}
    \psfrag{ 1}[c][c][1.2]{$10^{0}$}
    \psfrag{ 7}[c][c][1.2]{$7$}
    \psfrag{ 8}[c][c][1.2]{$8$}
    \psfrag{ 9}[c][c][1.2]{$9$}
    \psfrag{ 10}[c][c][1.2]{$10$}
    \psfrag{ 11}[c][c][1.2]{$11$}
    \psfrag{ 12}[c][c][1.2]{$12$}
    \psfrag{ 13}[c][c][1.2]{$13$}
    \psfrag{10}[c][c][0.8]{$\mathbf{10\%}$}
    \psfrag{30}[c][c][0.8]{$\mathbf{30\%}$}
    \psfrag{Optimistic}[c][c][1.0]{\textcolor{yellow}{\bf{Optimistic}}}
    \psfrag{Moderate}[c][c][1.0]{\textcolor{yellow}{\bf{Moderate}}}
    \psfrag{Pessimistic}[c][c][1.0]{\textcolor{yellow}{\bf{Pessimistic}}}
    \psfrag{ 0}[c][c][1.2]{$0$}
    \psfrag{ 10}[c][c][1.2]{$10$}
    \psfrag{ 20}[c][c][1.2]{$20$}
    \psfrag{ 30}[c][c][1.2]{$30$}
    \psfrag{ 40}[c][c][1.2]{$40$}
    \psfrag{ 50}[c][c][1.2]{$50$}
    \psfrag{ 60}[c][c][1.2]{$60$}
    \psfrag{ 80}[c][c][1.2]{$80$}
    \psfrag{ 100}[c][c][1.2]{$100$}
   \includegraphics[scale=0.7, angle=-90]{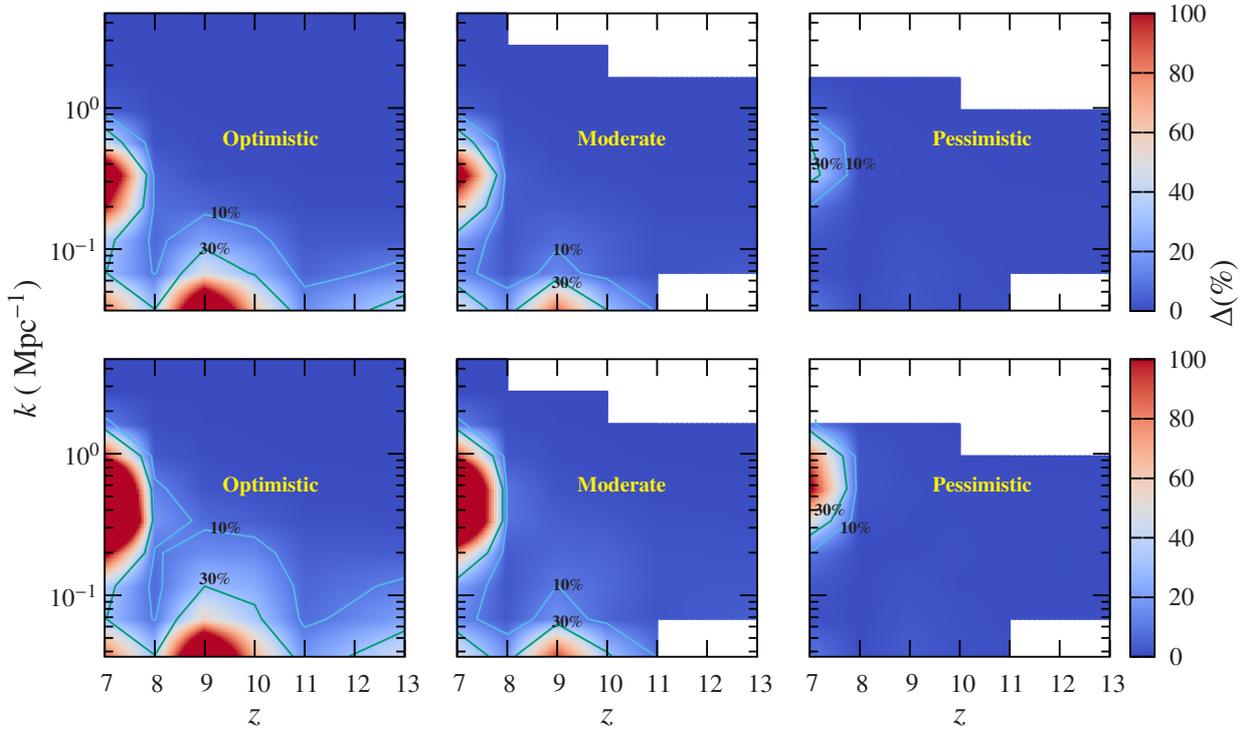}
    \caption{This shows $\Delta$ the percentage deviation of the non-Gaussian (Case--I) error predictions from the Gaussian predictions as a function of $k$ and $z$ considering the  three different foreground scenarios. The top and bottom panels show the results for $1024$ hours and $10000$ hours of observations, respectively.}
    \label{fig:frac_wedge}
\end{figure*}

Figure \ref{fig:frac_wedge} shows the percentage deviation $\Delta$  of the non-Gaussian error predictions (Case--I) with respect to the Gaussian predictions. Considering the Optimistic scenario discussed in the previous section (Figure \ref{fig:frac_1}), the deviations are prominent $(\Delta >  50 \%)$ at the smallest $k$ bin for $z=7$ and $9$ and also in the $k$ range $0.2\, -\, 0.5 \impc$ at $z=7$. The number of $k$ modes in each $k$ bin gets reduced due to the foreground wedge, and consequently the relative contribution to the error covariance (equation \ref{eq:e4})  from the trispectrum is reduced.  We therefore expect progressively smaller values of $\Delta$ as we go from the Optimistic to the Moderate and the Pessimistic scenarios. Considering the Moderate scenario, the results are similar to the Optimistic ones, however the values of $\Delta$ are somewhat smaller though they still exceed $50 \%$ (and $100 \%$ in some cases). For the Pessimistic scenario, however, the values of $\Delta$ are considerably smaller and they do not exceed $50 \%$ for $1024$ hours whereas they exceed $50 \%$ only  in the $k$ range $0.3 \, - \, 1  \impc $ at $z=7$ for $10000$ hours. 
\begin{figure*}
    \centering
    \psfrag{k}[c][b][1.5]{${k ~(\impc)}$}
   	\psfrag{k1}[c][c][1.5]{${k ~(\impc)}$}
   	\psfrag{r}[c][c][1.5]{$r_{ij}$}
   	\psfrag{1.0}[c][c][1.2]{$1.0$}
   	\psfrag{0.8}[c][c][1.2]{$0.8$}
   	\psfrag{0.6}[c][c][1.2]{$0.6$}
   	\psfrag{0.4}[c][c][1.2]{$0.4$}
   	\psfrag{0.2}[c][c][1.2]{$0.2$}
   	\psfrag{0.0}[c][c][1.2]{$0.0$}
   	\psfrag{-0.2}[c][c][1.2]{$-0.2$}
   	\psfrag{-0.4}[c][c][1.2]{$-0.4$}
   	\psfrag{-0.6}[c][c][1.2]{$-0.6$}
   	\psfrag{0.04}[c][c][1.2]{$0.04$}
   	\psfrag{0.07}[c][c][1.2]{$0.07$}
   	\psfrag{0.12}[c][c][1.2]{$0.12$}
   	\psfrag{0.20}[c][c][1.2]{$0.20$}
   	\psfrag{0.34}[c][c][1.2]{$0.34$}
   	\psfrag{0.57}[c][c][1.2]{$0.57$}
   	\psfrag{0.96}[c][c][1.2]{$0.96$}
   	\psfrag{1.63}[c][c][1.2]{$1.63$}
   	\psfrag{2.75}[c][c][1.2]{$2.75$}
   	\psfrag{4.66}[c][c][1.2]{$4.66$}
   	\psfrag{Optimistic}[c][c][1.0]{\bf{Optimistic}}
    \psfrag{Moderate}[c][c][1.0]{\bf{Moderate}}
    \psfrag{Pessimistic}[c][c][1.0]{\bf{Pessimistic}}
    \mbox{\includegraphics[width=0.32\textwidth,angle=-90]{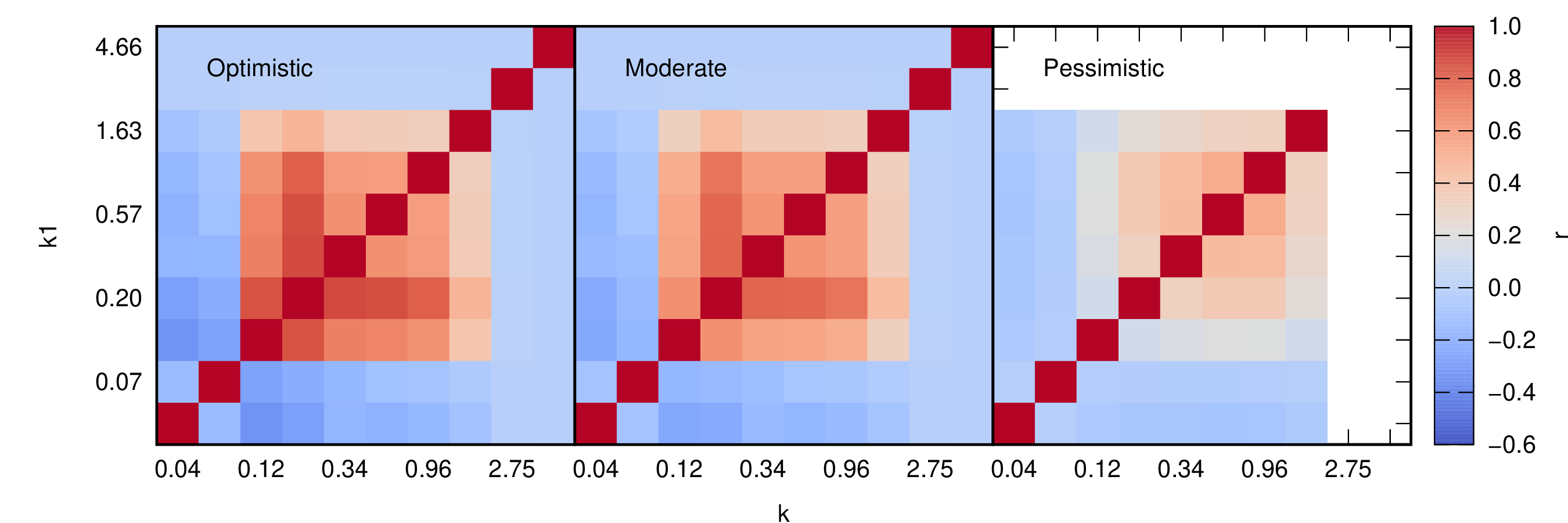}}
    \caption{This shows the correlation coefficients $r_{ij}$ at $z=7$ considering the three different foreground scenarios for $t_{\rm obs}=10000$ hours. In the Pessimistic scenario the  two largest $k$ bins are excluded  due to the Foreground wedge.}
    \label{fig:cov_wedge}
\end{figure*}

Figure \ref{fig:cov_wedge} shows the correlations between the different $k$ bins induced by the non-Gaussianity considering $10000$ hours. We have restricted the analysis to $z=7$, where we have prominent deviations from the Gaussian predictions for all the three scenarios. Comparing the Optimistic scenario with the lower left panel of Figure \ref{fig:cov_1024}, which shows the same for $t_{\rm obs}=1024$ hours we find that the extent of the positive correlation increases by one  $k$ bin and the values of the correlation coefficients $r_{ij}$ also increase. 
Comparing the left and centre panels of Figure  \ref{fig:cov_wedge}, we see that the pattern of correlations and anticorrelations has the same $k$ extent for the Optimistic and Moderate scenarios, however the magnitudes of $r_{ij}$ decrease by $10 \,- \,  30 \% $. Considering the Pessimistic scenario, we find that the anticorrelation between the  two smallest $k$ bins and the larger $k$ bins is not noticeable here.  The extent of the $k$ bins with positive correlations is the same as the Optimistic scenario, but the values of $r_{ij}$ are  $60 \, - \, 70 \% $ smaller. Considering other redshifts for which the results are not shown here, we find that there are some correlations between the different $k$ bins also at $z=9$ in the Moderate scenario, however these are absent in the Pessimistic scenario. These correlations for the Moderate scenario are however considerably smaller and they are $\sim 50 \%$ of the correlations seen in the bottom-right panel of Figure \ref{fig:cov_1024} .

Summarizing  this section, we find that foregrounds restrict the $k$ modes that can be used for detecting the EoR 21-cm PS. This results in reducing the SNR and also reducing the impact of non-Gaussianity on the error predictions. The deviations from the Gaussian predictions continue  to be important ($> 50 \%$) at $z=7$ even if the effect of Foreground Avoidance is included.
 

\section{Summary and Conclusions}\label{sec:dis}
There are currently several radio-interferometric arrays such as LOFAR, MWA and  PAPER which have been carrying out observations to detect the EoR 21-cm PS. Several other instruments like  HERA and SKA, which are expected to have greater sensitivity, are under construction or planning.  It is of considerable interest to have error predictions for the EoR 21-cm PS considering such observations, and there have been several works (e.g. \citealt{Mellema_2013,Pober_2014,21cmmc,Ewall-Wice}) addressing this  under the assumption that the EoR 21-cm signal is a Gaussian random field. However there have been several studies (e.g. \citealt{Pandey_2005,Mondal_2015,Mondal_I,Mondal_II,Majumdar_2018}) that show that the EoR 21-cm signal is non-Gaussian in nature. In this paper we study how these non-Gaussianties affect the error estimates for the EoR 21-cm PS considering observations with the upcoming SKA-Low.

The error predictions for any observation of the EoR 21-cm PS are quantified through the error covariance matrix $\cov_{ij}$, which depends on the PS and  the trispectrum of the EoR 21-cm signal, and also observational effects like the array baseline distribution and the system noise. The EoR simulations generally provide predictions for the bin-averaged 21-cm PS and  trispectrum without incorporating the observational effects. In this paper we first present a methodology for calculating $\cov_{ij}$ combining the simulated PS and trispectrum with these  observational effects.  The error covariance matrix for the binned 21-cm PS (equation \ref{eq:cov})
actually depends on the trispectrum $T(\kk_{g_{i}}, -\kk_{g_{i}}, \kk_{g_{j}},-\kk_{g_{j}})$ evaluated at individual pairs of Fourier modes  $\kk_{g_{i}}$ and   $\kk_{g_{j}}$, unfortunately this is not available from simulations  as the computations involved for a reliable estimate is extremely large and cumbersome. We have overcome this by considering two different cases where we approximate $T(\kk_{g_{i}}, -\kk_{g_{i}}, \kk_{g_{j}},-\kk_{g_{j}})$  using  the bin averaged trispectrum   $\bar{T}(k_i,k_j)$
 for which estimates are available from simulations.
Results are mainly presented  for Case--I which assumes that  the  different  $\kk$ modes within the same $k$  bin are  completely correlated. We also consider Case--II which assumes the  different  $\kk$ modes within the same $k$  bin to be totally uncorrelated.  These represent two extreme cases, and the reality is expected to be somewhere in between. We find that the error predictions for  Case--II are typically intermediate between  the Gaussian predictions and Case--I.  In most situations  we may adopt a simple picture where the predictions for Case--I  represent the upper limit for the error covariance matrix, and the actual errors may be expected to have  values between these and the Gaussian predictions. It may however be noted that we do have a few  situations where the predictions for Case--II exceed those for Case--I  as seen in the lower left-hand panel of Figure \ref{fig:SNR_8}.

We find that the predicted  errors typically increase at large $k$ (Figure \ref{fig:kkd_1}) where it is system noise dominated. In this situation the  r.m.s. error scales  as $t_{\rm obs}^{-1}$, and  the $k$ range below which a $5 \sigma$ detection of the EoR 21-cm PS is  possible  ($k_m$) increases as $t_{\rm  obs}$ is increased (Figure \ref{fig:km}). The values of $k_m$ also increase as reionization proceeds as $T_{\rm sys}$  increases  with redshift. At all $z$ a $5 \sigma$ detection is possible for $128$ hours of observation. However  $k_m$ is largest $(\sim 0.9  \impc)$ at $z=8$, and the accessible $k$ range is smaller at higher $z$ with $k_m\sim 0.09  \impc$ at $z=13$. The value of $k_m$ increases significantly for $t_{\rm obs}=1024$ hours and we have $k_m > 1  \impc$ for all $z \le 10$. We have $k_m > 1  \impc$ at all redshifts for $t_{\rm obs}=10000$ hours. We note that at redshifts $z=7$ and $9$ a $5 \sigma$ detection is not possible at the smallest $k$ bin $(k=0.04 \impc)$, which is predicted to be  cosmic variance limited (Figures \ref{fig:kkd_1} and \ref{fig:SNR_1}). 

The error predictions here are in excess of the Gaussian predictions that ignore the contribution from the trispectrum. At all $z$  the fractional deviation $\Delta$ is found to exhibit a `U' shaped $k$ dependence (Figure \ref{fig:frac_1})  in the CV limit where the system noise can be ignored. The deviations are minimum at $k_{\rm min} \sim 0.1-0.3 \impc$ where the ratio $\Nk \bar{T}(k_i,k_i)/\bar{P}^2(k_i)$ also is minimum, and $\Delta$ rises steeply on both sides with particularly large values $(\sim 100 \%$) at $k > k_{\rm min}$. For finite observation times where the system noise is important, we have significant deviations ($\Delta \sim 40 - 100 \%$) at $k < k_{\rm min}$ for $t_{\rm obs}=1024$. However, for $k> k_{\rm min}$ the errors are system noise dominated (except at $z \le 8$) and the deviations are small. At $z=7$ we have particularly large deviations ($\Delta \sim 100 \%$ and larger) at $k> k_{\rm min}$ for $t_{\rm obs} \ge 1024$ hours.

The SNR (Figures \ref{fig:SNR_6} and \ref{fig:SNR_8}) is expected to increase $\propto t_{\rm obs}$ for small observation time where the system noise dominates the errors; we also expect the Gaussian predictions to match those for Case--I and Case--II in this regime. This is clearly seen for most redshifts at $k=0.57 \impc$ (Figure \ref{fig:SNR_6}) and $1.63 \impc$ (Figure \ref{fig:SNR_8}), which are, respectively, representative of intermediate and small length-scales. However, at $z=7$ we see that the SNR saturates at the CV limit beyond $t_{\rm obs}\sim 10^3$ hours. At $k=0.04 \impc$ (Figure \ref{fig:SNR_1}), which is representative of large length-scales, the SNR saturates within $\sim 100$ hours at all redshifts. The Gaussian predictions, Case--I and Case--II, also  differ significantly, and the predictions for Case--II are typically between the Gaussian and Case--I predictions.

The inherent non-Gaussianity of the EoR 21-cm signal introduces correlations between the errors in different $k$ bins. Although $\cov_{ij}$ ($i\neq j$) is independent of $t_{\rm obs}$, the dimensionless correlation coefficients $r_{ij}=\cov_{ij}/\sqrt{\cov_{ii} \cov_{jj}}$ are $t_{\rm obs}$ dependent. We expect the correlations $r_{ij}$ to become important for large $t_{\rm obs}$, and we have presented results for $1024$ hours (Figure \ref{fig:cov_1024}). We find significant correlations and anticorrelations $\mid r_{ij} \mid \sim 0.1-0.4$ among the four smallest $k$ bins over the entire $z$ range. Further, we find strong correlations $r_{ij} \sim 0.7-0.8$ among some of the $k$ bins in the range $k \sim 0.1 - 1 \impc$ at $z=7$ and $9$.

The results summarized  till now has not considered the foregrounds. The foreground contamination is expected to be restricted within a wedge, and only the $k$ modes outside this foreground wedge can be used for 21-cm PS detection. In addition to the Optimistic scenario where there are no foregrounds, we have also considered the Moderate and Pessimistic scenarios where the $(\kk_{\perp},k_{\parallel})$ extent of the foreground wedge respectively correspond to  $\theta_{\rm L}=3\times {\rm FWHM}/2$ and $\theta_{\rm L}=90^{\circ}$ in equation (\ref{eq:wedge}). We find that for both the foreground scenarios a few $k$ bins are excluded 
and the SNR is reduced compared to the Optimistic scenario  (Figure \ref{fig:SNR_wedge}). The impact of non-Gaussianity on the error predictions is also reduced (Figure \ref{fig:frac_wedge}). The results 
for the Moderate scenario are comparable to those for the  Optimistic scenarios, which have no foregrounds, however the predictions are considerably degraded for the Pessimistic scenario. Finally we note that the deviations from the Gaussian predictions, including correlations between the different $k$ bins, continue  to be important ($> 50 \%$) for all the scenarios at $z=7$. 

In conclusion, we note that non-Gaussian effects make a significant contribution to the error predictions, particularly at low redshifts and large length-scales. In addition to increasing the error predictions with respect to the Gaussian predictions, it also introduces significant correlations and anticorrelations between different $k$ bins.


\section*{Acknowledgments}
The authors would like to thank Dr. Raghunath Ghara and Srijita Pal for the help related to the specifications of SKA-Low and baseline distributions. AKS would like to thank Dr. Anjan K. Sarkar, Debanjan Sarkar and Suman Chatterjee for the fruitful discussions. RM would like to acknowledge funding form the Science and Technology Facilities Council (grant numbers ST/F002858/1 and ST/I000976/1) and the Southeast Physics Network~(SEPNet).
\bibliography{ref}
\label{lastpage}
\bsp	
\end{document}